\documentclass{article}
\usepackage{amsmath}
\usepackage{amssymb}
\usepackage{authblk} 
\usepackage{graphicx}
\usepackage{hyperref}
\usepackage{url}
\usepackage{pbox}
\usepackage{longtable}
\usepackage{booktabs}
\usepackage{array}
\usepackage{adjustbox}
\usepackage{enumerate}
\usepackage{float}
\usepackage{xcolor}
\usepackage{bm}
\usepackage[margin=1in]{geometry}
\usepackage{enumitem}
\usepackage{array}
\newcolumntype{L}[1]{>{\raggedright\arraybackslash}p{#1}}
\usepackage[table]{xcolor}
\usepackage[authoryear]{natbib}
\usepackage{boxedminipage}
\usepackage[framemethod=pgf]{mdframed}
\usepackage{mdframed,adjustbox,float}
\usepackage{xcolor}

\usepackage{float}
\usepackage{caption}

\definecolor{myorange}{HTML}{FF8C00} 
\definecolor{darkblue}{rgb}{0,0,0.5}
\definecolor{mypurple}{HTML}{6A0DAD} 
\hypersetup{
    colorlinks=true,
    linkcolor=darkblue,
    citecolor=darkblue,
    urlcolor=darkblue,
    filecolor=black
}

\makeatletter
\renewcommand\AB@affillist{}
\makeatother

\title{\vspace{-1cm}\textbf{AI and jobs. \\\Large A review of theory, estimates, and evidence}\thanks{The authors would like to thank Leonard Bocquet, University of Cambridge, for helpful research assistance. We would also like to thank Piotr Lewandowski and Qin Chen for sharing their exposure scores with us.}\\[1ex]}

\author[1]{R. Maria del Rio-Chanona}
\author[2]{Ekkehard Ernst}
\author[2]{Rossana Merola}
\author[2]{Daniel Samaan}
\author[3]{Ole Teutloff}
\affil[1]{Department of Computer Science and AI Centre, University College London}
\affil[2]{International Labour Organisation (ILO)}
\affil[3]{Oxford Internet Institute, University of Oxford; Copenhagen Center For Social Data Science, University of Copenhagen}
\date{\textbf{Pre-print} — \today}

\begin{document}
\pagenumbering{gobble}

\maketitle
\vspace{-8ex}

\begingroup
\renewcommand\thefootnote{}\footnote{Disclaimer:  Any view expressed or conclusions drawn in this discussion paper represent the views of the authors and do not necessarily represent ILO views or ILO policy. The views expressed herein should be attributed to the authors and not to the ILO, its management or its constituents.}
\addtocounter{footnote}{-1}
\endgroup

\begin{abstract}
Generative AI (GenAI) is altering work processes, task composition, and organizational design, yet its effects on employment and the macroeconomy remain unresolved. In this review, we synthesize theory and
empirical evidence at three levels. First, we trace the evolution from aggregate production frameworks to
task- and expertise-based models and organizational perspectives. Second, we quantitatively review and
compare (ex-ante) AI exposure measures on jobs and occupations from multiple studies and find convergence
towards high-wage jobs. Third, we assemble ex-post evidence of AI's impact on employment from randomized controlled trials (RCTs), field experiments, and digital trace data (e.g., online labor platforms, software repositories), complemented by partial coverage of surveys and vacancy studies. To better differentiate the reviewed studies by their observed outcomes, we introduce a classification that distinguishes simple from complex tasks using four dimensions: knowledge, clarity of goal, interdependence, and context requirements. Across the reviewed studies, productivity gains are sizable but context-dependent—on the order of $\sim$20--60\% in controlled RCTs, and 15-30\% in field experiments. Novice workers tend to benefit more from LLMs in simple tasks. Across complex tasks, evidence is mixed on whether low or high-skilled workers benefit more. Digital trace data show substitution between humans and machines in writing/translation alongside rising demand for AI, with mild evidence of declining demand for novice workers. A more substantial decrease in demand for novice jobs across AI complementary work emerges from recent studies using surveys, platform payment records, or administrative data. Research gaps include the focus on simple tasks in experiments, the limited diversity of LLMs studied, and technology-centric AI exposure measures that overlook adoption dynamics and whether exposure translates into substitution, productivity gains, erode or increase expertise.
\end{abstract}

\newpage
\adjustbox{pagecenter, varwidth=\textwidth,scale=1}{
\tableofcontents
}
\newpage
\section*{Summary of key findings}
\addcontentsline{toc}{section}{Summary of key findings}

\setlength{\intextsep}{6pt}   
\setlength{\floatsep}{6pt}    

\captionsetup[figure]{aboveskip=3pt, belowskip=0pt}
\pagenumbering{roman}

\renewcommand{\figurename}{Box}


\begin{figure}[H]
\caption{\label{box:theoryExposure}Theory and exposure measures}
\adjustbox{pagecenter, varwidth=\textwidth,scale=0.99}{
\begin{mdframed}[linecolor=blue,outerlinewidth=2,roundcorner=7pt]
\footnotesize

\begin{enumerate}

\item\textbf{Technology’s impact on job creation or destruction depends on the extent to which productivity gains boost consumer demand, the extent to which it automates work, and the emergence of new tasks.}
 When a technology can easily substitute for human workers, it can trigger employment shifts away from people. However, technology-driven productivity gains reduce prices and can boost employment when consumers respond by buying much more, either in the current or other sectors (reallocating demand). Technologies may also create entirely new occupations that can offset displacement effects. 

\item \textbf{Technology can complement workers by targeting a subset of their tasks, in this case, wage effects depend on whether it democratizes or emphasizes expertise.} Technology can increase wages when it handles tasks that require less expertise and expands demand for human specialists, or decrease wages when it makes expert skills accessible to more people and expands the pool of qualified applicants and employment. 

\item \textbf{Tasks can be classified as simple or complex using four dimensions.} Building on previous work on routine and easy-to-learn tasks as well as collective intelligence, we introduce a classification that distinguishes simple from complex tasks using four dimensions: knowledge requirements, clarity of goals, task interdependence, and context requirements. Tasks scoring high on at least three dimensions are classified as complex.


\item \textbf{AI exposure correlates with high wages but varies by geography and gender, with unclear implications for workers.} Occupations with higher wages show greater AI exposure. High-income countries show 5.5\% of jobs at high AI exposure versus 0.4\% in low-income countries, while women face 8.5\% exposure versus 3.9\% for men in high-income countries. Job mobility across occupations means net exposure rates might be larger than gross ones for occupations tightly connected to highly affected ones. Whether exposure is beneficial or detrimental for workers remains unclear.

\end{enumerate}

\end{mdframed}}
\end{figure}

\begin{figure}[H]
\caption{\label{box:empirical}Empirical evidence}
\adjustbox{pagecenter, varwidth=\textwidth,scale=0.99}{
\begin{mdframed}[linecolor=myorange,outerlinewidth=2,roundcorner=7pt]
\footnotesize

\begin{enumerate}

 \item \textbf{Productivity gains are significant but context-dependent.} Randomized Controlled Trials (RCTs) show productivity increases of 20–60\%, in field experiments, between  15-30\% in field experiments, and generally lower gains in natural experiments, indicating implementation and integration challenges. Teams between humans and AI also show significant productivity gains such as idea generation, marketing ads or coding (section \ref{sec:team-performance}).

\item \textbf{The impact on inequality is inconclusive.} In controlled experiments, the use of AI for simple tasks tends to benefit primarily inexperienced and less well-performing users. However, these gains may reflect "passive pasting"—workers submitting AI output without modification, pointing towards automation rather than learning. For complex tasks or in field experiments, the impact is mixed and depends on the particular task at hand and the setting under which the introduction of AI was observed (sections \ref{subsec:controlled_experiments} and \ref{subsec:field_experiments}).

\item \textbf{Demand shifts from substitutable to complementary skills, with new AI-specific roles emerging in China and online labor markets.} In China, job postings shifted toward broader complementary skills like creativity and problem-solving, while easily substitutable skills like documentation and design declined. Online labor markets show 20-50\% demand decreases for substitutable skills like writing and translation, while AI-specific tasks increased in demand. 

\item \textbf{Reduced demand for novice workers in the U.S. and online labor markets; not in Denmark.} U.S. companies adopting AI reduced hiring of junior employees (ages 22-25) by 13\%, with mid-tier graduates hit hardest, while online labor markets show decreased demand for novice workers in AI complementary roles. However, in Denmark, economic effects remain negligible with impacts smaller than 1\%. 

\item \textbf{Context, relationships, and metacognition shape AI productivity gains for workers and teams.}  Workers achieve greater improvements when they possess project-specific familiarity and accurate self-assessment of their abilities. Teams benefit more from AI when members have higher cognitive ability, stronger internal relationships, and larger sizes.

\item \textbf{Organizational inertia limits transformation.} While dynamic environments like open-source software show restructuring (e.g., developers focusing more on coding and less on management), traditional corporate settings often see minimal organizational change despite individual productivity gains (section \ref{subsec:nat_experiments}).

\item \textbf{AI can reduce coordination overhead.} Evidence from software development shows that AI adoption leads to a reallocation of time from project management and coordination activities towards core technical work, making work more autonomous (section \ref{subsec:teamsTakeAways}).

\item \textbf{AI introduces a trade-off between individual vs collective creativity.} Across different fields AI enhances individual creative performance but reduces collective diversity. Yet there also positive effects AI increases teams productivity and bridge disciplinary silos (section \ref{subsec:teams_creativity}).

\end{enumerate}

\end{mdframed}}
\end{figure}

\begin{figure}[H]
\caption{\label{box:researchGaps}Research gaps}
\adjustbox{pagecenter, varwidth=\textwidth,scale=0.99}{
\begin{mdframed}[linecolor=mypurple,outerlinewidth=2,roundcorner=7pt]
\footnotesize

\begin{enumerate}

\item \textbf{Comparative statics struggle to model labor market adjustment dynamics.} Most of current theoretical frameworks treat technological change through comparative statics, comparing pre- and post-shock equilibria. In reality, displaced workers face unemployment spells, skill mismatches, and wage rigidities that make transitioning between jobs costly and slow.  

\item \textbf{The experimental literature focuses primarily on simple tasks.} Over 70\% of experimental studies focus on simple, well-defined tasks, creating a significant research gap regarding AI's impact on the complex, ambiguous, and interdependent work common in professional settings.

\item \textbf{Exposure measures are technologically deterministic.} Prevailing metrics assess the \textit{feasibility} of automating tasks but ignore critical economic, legal, ethical, and social factors that ultimately determine the \textit{likelihood and speed} of adoption.

\item \textbf{The experimental literature shows limited AI model diversity.} Most studies employ GPT-based models (including Copilot) or specialized applications for specific domains. Few studies examine other frontier models like Claude, Gemini, or DeepSeek.

\end{enumerate}

\end{mdframed}}
\end{figure}

\newpage 
\pagenumbering{arabic}

\section{Introduction}

Recent advances in generative AI (“GenAI”) have reignited debates over large-scale job losses from digital technologies.\footnote{Dario Amodei, for instance, the CEO of Anthrophic, a leading digital company developing the large language model "Claude.AI" recently warned about a rapid rise in unemployment as a significant share of entry-level jobs might disappear, thereby preventing especially young people from entering succesfully the labour market \url{https://www.axios.com/2025/05/28/ai-jobs-white-collar-unemployment-anthropic}.}  Earlier waves of AI tools in the 2010s provoked similar concerns, yet widespread employment effects have not materialized to date. With the pace of progress now seemingly accelerating, however, many observers expect a far more disruptive impact on labour markets in the near future.

These fears have been triggered by the significant shift in scale, scope and performance artificial intelligence went through over the past decade and a half. The arrival of chatbots such as Apple's Siri marked the beginning of interactive tools open for the general public. Thanks to the abundance of big data collected through widely available smart-phones, increasing computing power, and the continuous improvement of methods to train statistical models for pattern recognition, a large array of tools have been quickly integrated into the existing digital economy, for example into e-payment services, online warehouses or social media. Online labour platforms have become an important channel through which micro-tasks are mediated globally, including AI-related activities such as data labelling as well as new digital occupations like content creation. These developments have reshaped existing work arrangements but also generated additional opportunities for income \citep{ilo21}. The latest large language models (LLMs), capable of generating coherent text in response to human-language instructions, have further accelerated the adoption of AI tools in both professional and personal contexts. Currently, almost 40 per cent of US-Americans are is using one of the various applications that run under the heading "Generative AI", whether it is for text, sound, image or video production\footnote{\url{https://www.stlouisfed.org/on-the-economy/2024/sep/rapid-adoption-generative-ai}}. To date, however, the labour market implications remain hotly debated. Moreover, given the recency of the latest generation of AI tools, data that would show any significant trend remains scarce. Instead, a large and rapidly growing literature focuses on ex-ante assessments of the automation potential of (Gen)AI as well as on occupation-specific experimental evidence of the use of certain tools for specific tasks. 

In this paper, we review the current state of evidence and knowledge on the potential for job losses and transformation following the development and widespread adoption of the latest wave of AI tools. There are several features that characterize the studies we review here: Most  focus on job loss and transformation, while few assess the potential for job creation, and—according to our knowledge — no academic work attempts to systematically predict new occupations (though some policy reports occasionally speculate about them). This makes it difficult to provide a net assessment on future developments of aggregate unemployment. Second, most of the studies we review analyze the impact of AI from a task-based perspective, assessing only the technological feasibility of labour substitution by machines. Third, only few papers  assess also the potential or actual adoption of AI tools, which would, however, be necessary to evaluate the speed and scope of changes induced by AI.

Our literature review expands on similar efforts by \citep{cruces24,Cazzaniga2024} and others in that it provides a systematic overview of the key methods used to identify labour market effects of GenAI, their strengths and weaknesses. In particular, it opens the perspective to a range of approaches that look at how (Gen)AI might help automate entire functions in companies, thereby significantly raising not only individual performance but that of teams and corporations altogether. Our paper focuses on the most recent wave of papers published around the impact of GenAI on jobs. Sporadically, we refer to papers released after the previous wave of AI development published during the 2010s. While no broad consensus emerges (yet) out of these various papers, our review allows us to draw a few general observations (see boxes \ref{box:theoryExposure}-\ref{box:researchGaps}).

\renewcommand{\figurename}{Figure}
\setcounter{figure}{0}

The paper is structured as follows: Section 2 reviews the conceptual frameworks in economics that are used to analyze the effects of automation and AI on the labor market, including the production function approach, task-based models, and team-oriented perspectives. Section 3 presents and discusses current empirical estimates of AI exposure across jobs and occupations, which continue to dominate academic and policy debates. We critically assess the strengths and limitations of these measures and review convergence in the literature regarding which jobs, activities, or occupations are more exposed to AI, and whether this exposure implies replacement or transformation of work. Section 4 examines the most recent ex-post evidence on the labor market effects of generative AI. We summarize findings from randomized controlled trials, field experiments, natural experiments, and large-scale digital trace data, highlighting their implications for productivity, job substitution, and work organization. Section 5 extends the analysis beyond jobs to collective performance, covering areas such as creativity, innovation, marketing, coding, and algorithmic management. Finally, Section 6 concludes and identifies gaps for future research.

\section{Understanding the impact of automation on jobs\label{sec:theory}}

To assess the labour market impact of AI, two main approaches exist in the literature: technological change  modeled through the macroeconomic production function in a growth theoretical approach, first popularized by \cite{solow56} and the the task framework \cite{Autor2013}. In this section, we review both main conceptual approaches and discuss their implications for understanding the impact of AI on jobs.

\subsection{Growth theory and production functions}

The traditional approach to understanding technological change's impact on labor markets comes from macroeconomic growth theory, pioneered by \citep{solow56}). This framework treats the economy as an aggregate production function combining capital and labor, with technological progress shifting this function over time. While originally developed to explain long-run growth patterns, these models can be applied to understand AI's potential effects.



The original Solow model made use of a Cobb-Douglas production function with homogenous factor inputs from capital and labour, with a unit-elasticity of substitution between the two production factors. In this case, the only form of technological change that entails a balanced growth path for the economy requires labour-augmenting technological change (Harrod-neutral technological change).\footnote{This was formally proved by Uzawa in 1961.\url{https://en.wikipedia.org/wiki/Uzawa\%27s_theorem}} In this case, technological improvements yield continuous increases in returns to both capital and labour, without generating unemployment or inequality. Any improvements in productivity will immediately translate into an increase in real wages (assuming full employment) thereby keeping the distribution of income between capital and labour constant. In equilibrium, therefore, neither unemployment (by assumption) nor changes in income inequality is observed in this set-up.

However, the model proved too restrictive to explain declining labor shares observed across advanced economies since the 1980s \citep{Karabarbounis14}. This issue resulted because, by construction, the Cobb-Douglas framework ruled out that technological change might benefit capital at labor's expense.\footnote{
In the Cobb-Douglas production function, $Y=AK^\alpha L^{1-\alpha}$, the exponents dictate the income shares. By mathematical derivation, labor's share is always fixed at $1-\alpha$ and capital's share at $\alpha$. This is a direct consequence of the function's unitary elasticity of substitution ($\sigma=1$), which prevents technology ($A$) from altering this distribution} The solution was to 
relax the assumption of unit elasticity of substitution, which is, in reality, rarely satisfied (see \cite{rowthorn99}). The constant elasticity of substitution (CES) production function, addressed these limitations by allowing flexible substitution between capital and labor. This framework revealed that technological change can be explicitly \textit{biased}, benefiting one factor while potentially harming the other.

How easily firms can substitute between workers and machines determines the impact of technology on employment and wages. When capital and labor are highly substitutable (like self-checkout machines replacing cashiers), even small improvements in technology can trigger large shifts away from human workers. Conversely, when they are complementary (like pilots working with autopilot systems), technological improvements make human workers more valuable. The CES framework showed that labor-augmenting technological progress, i.e., making workers more efficient, can paradoxically reduce employment when strong complementarity limits firms' ability to expand output \citep{AcemogluRestrepo2019}. A large literature has focused since on measuring the capital-labor elasticity of substitution \citep{Karabarbounis14, oberfield21} \footnote{The estimates vary widely across studies (\cite{Karabarbounis14} find epsilon $> 1$ whereas \cite{oberfield21} find epsilon $< 1$), stemming in large parts from methodological differences and identifying assumptions. In particular, \cite{Karabarbounis14} assume technological shifters A and B are orthogonal to price changes, while \cite{oberfield21} don't. At the macro-economic level the capital-labor elasticity might differ substantially from the micro-level one. For instance, \cite{houthakker55} shows how Leontief production functions (epsilon = 0) might be aggregated into a Cobb-Douglas production function (epsilon = 1). In other words, this macro elasticity would not informative about the sectoral-level or firm-level AI-labor substitution, which is ultimately what matters when predicting the labor market effects of AI. Here, $K$ aggregates too many different capital types to be specifically informative about AI. Hence the importance of accounting for various capital types (and labor types). \cite{oberfield21, baqaee20} among others show how to aggregate these micro-elasticities into macro-elasticities.}

The CES-production function is the backdrop to approaches by \cite{aghion19} and \cite{korinek24, korinek25} to understand how an (exogenously given) increase in the share of automatable jobs driven by AI will impact the capital-labour ratio and the labour income share. As in the canonical Solow growth model, capital investment is determined by the savings rate whereas wage growth depends on the speed of automation. Provided technological progress is not too fast, the labour income share can remain stable in this model even if the share of automatable jobs increases. The wage share will only collapse under the assumption of "Artificial General Intelligence" where all jobs can eventually be automated \citep{korinek24, korinek25}.

Following the earlier literature on endogenous growth models, \cite{aghion19} extend their models to distinguish between production workers and those contributing to increase the share of automatable jobs. In this case, eventually all workers will switch to become "researchers", which will contribute to an accelerating growth rate and a collapse of the capital income share. Automating even the research function then leads to a "singularity", i.e., a situation where economic growth explodes \citep{aghion19}.

None of these approaches using the production function approach consider the demand side. However, as demonstrated by \cite{Bessen2018,bessen19}, sectoral shifts in the demand for labour will depend on the relative Engel's curves of each sector, i.e. the elasticity of sector demand with changes in the household income: As households become richer, they might not necessarily demand more cars or fridges - so job growth in these sectors will decline in line with productivity growth - but are likely to demand more health care services where the income elasticity is larger than one. In other words, when productivity growth leads to increased household incomes, it might increase job growth but not uniformly, leading to job destruction in some sectors (and regions) and job growth in others with a net positive outcome for employment. In this respect, to the extent that AI-induced automation affects particularly inelastic sectors, job growth and inequality could be affected more strongly than suggested by the canonical growth model.

Moreover, wage inequality did not only rise between sectors but also between jobs in the same sector, a further empirical puzzle that the original Solow-growth model did not consider. Indeed, with the advent of the computer revolution aggregate productivity was boosted at the same time as wage gaps widened in the US.  By treating labor as a single input, the canonical production function approach was silent on why some workers benefited from new technology while others were left behind. Multi-worker production functions addressed this puzzle by distinguishing between worker types, typically high-skill and low-skill labor \citep{katz92,autor98}. These models revealed two crucial mechanisms. First, skill-biased technological change: computers disproportionately enhanced the productivity of highly educated workers. Second, capital-skill complementarity: educated workers were strong complements to sophisticated new capital, while less-educated workers were often substitutes \citep{krusell00}.

\cite{caunedo23} extend the CES framework by estimating how easily capital substitutes for labor in different occupations, using data that links specific machines to the jobs that use them (e.g., industrial robots in manufacturing, diagnostic equipment in healthcare).Their model explains reallocation patterns from 1980 to 2015—such as the decline of typists versus the growth of other clerical workers, or the expansion of nurse practitioners versus the decline of some medical technicians—patterns uniform-elasticity models miss.

However, this approach treats substitution patterns as fixed technological parameters rather than outcomes of work organization. The same technology may replace workers in one context but complement them in another, depending on how firms structure tasks. Occupations themselves also evolve with technology: bank tellers shifted from cash handling to customer service, radiologists from film reading to intervention planning. These shifts in tasks, not just workers, cannot be captured by production-function models. This motivates the task-based framework discussed in the next section.



\subsection{Breaking up occupations: The task framework\label{sec:task-approach}}

Aggregate production functions mask how technologies affect different jobs. The task-based framework opens this ``black box'' by decomposing production into tasks performed by either labour or capital, allowing direct analysis of substitution and complementarity. Pioneered by \cite{zeira98} and developed by \cite{acemoglu11, acemogluRestrepo18, AcemogluRestrepo2019, acemogluRestrepo20, acemoglu24}, this framework now anchors analyses of AI's labour-market effects. 
Importantly, the theoretical insights from this framework are supported by empirical data on job content. Databases such as O*NET in the United States provide descriptions of occupational tasks, allowing to map theoretical predictions to observable changes in the labour market; see Section~\ref{sec:AI-jobs}.


The task production model assumes firms choose (i) capital, (ii) labour, and (iii) the assignment of tasks to each \citep{acemogluRestrepo22, acemogluKongRestrepo24}. \emph{Within a given task}, labour and capital are modeled as (often perfect) substitutes—the firm picks the cheaper method. \emph{Across tasks}, production is complementary (Leontief): all required tasks must be completed to produce a unit. Think of a checklist on an assembly line—welding the chassis does not substitute for installing brakes; skipping either yields no finished car.\footnote{A common simplification is perfect substitution between labour and capital \emph{at the task level} (each task is done entirely by one input). \cite{acemogluRestrepo18} show that allowing mixed, capital- and labour-intensive methods leaves the core results intact.} Firms therefore assign each task $t$ to the cheaper input by comparing unit costs. As AI improves or capital prices fall, more tasks flip to capital, reorganising production.

Firms endogenously allocate tasks to either labour or capital. For each task $t$, firms compare the unit cost of production using labour versus capital and assign the task to the cheaper input. This endogenous allocation makes the framework well suited to study automation: as AI improves or capital prices fall, tasks shift from labour to capital and production reorganises. 

\cite{acemogluRestrepo22} show that this task-based approach can be mathematically expressed as a CES production function where firms use labor and capital to produce output. Building on the framework discussed earlier, the key insight is that unlike traditional models where factor shares are fixed parameters, here the endogenous task shares—the proportion of tasks performed by workers versus machines—become flexible and respond to technological and cost changes. When technology advances or costs shift, firms don't just hire more or fewer workers—they fundamentally reorganize which tasks humans versus machines handle. This flexibility captures a crucial real-world phenomenon: technological change reshapes not just how much labor firms demand, but how production itself is structured.

This reorganization occurs through two distinct channels. \emph{Intensive margin effects} improve how efficiently workers or machines perform their existing tasks without changing who does what. A better software tool that helps an accountant work faster exemplifies this—the accountant still handles the same tasks, just more efficiently. However, the distributional effects depend on substitution patterns: labor-augmenting improvements can paradoxically reduce employment when workers and machines are strong complements, limiting firms' ability to expand output proportionally.

\emph{Extensive margin effects}, by contrast, shift tasks between humans and machines through two mechanisms. (i)Automation moves existing tasks from workers to capital—like AI systems now handling customer service inquiries previously managed by human agents. This creates competing forces: displacement as fewer workers are needed for these tasks, versus productivity gains as output increases. \cite{acemogluKongRestrepo24} warn of "so-so automation" where machines become only marginally better than humans at certain tasks. While firms may still automate to cut costs, the productivity gains prove modest, leaving displaced workers worse off without substantial economic benefits.
New task creation, the second extensive-margin channel, works in the opposite direction by generating entirely new jobs for workers—think of app developers, data scientists, or drone pilots. These reinstatement effects can offset automation's displacement, potentially expanding employment even as some traditional tasks disappear.

To account for the unequal impact of new technologies on occupational employment, \cite{autor03, autor25} expanded the task framework to classify different tasks regarding certain characteristics that would make them more or less amenable for automation. Specifically, the authors distinguished between routine- and non-routine tasks as well as between those tasks that require a certain level of expertise and others that don't. Routine tasks are those that can be summarised in programmed rules and executed by machines following explicit instructions. Similarly, expert tasks are those requiring certain (formal or informal) levels of expertise, such as training certificates or licensing requirements to be executed by employees.

Moreover, manual tasks can be distinguished from cognitive ones, depending on whether a tasks involved primarily physical activities, coordination, and dexterity versus those requiring mental capacity for analysis, communication and decision making. However, this distinction seems to be insufficient to understand the differential impact of GenAI on occupations: Previously non-routine cognitive tasks, for instance, are now being subject to automation (i.e., become "routine") thanks to pattern recognition at scale and the possibility for AI models to follow implicit rules learned from vast datasets \citep{ernst19}. Rather, a distinction needs to be made between simple and complex tasks to better characterise GenAI's labour market impact (see below and section \ref{subsec:ex-post}).

In this respect, the distinction between expert- and non-expert tasks offers a possibility to better account for task complexity and how GenAI can impact job opportunities \citep{autor25}. Specifically, those applications of AI that reduce the required level of expertise can enhance the pool of applicants for a specific job, thereby driving down the wage premium. On the other hand, when applications reduce the number of non-expert tasks to be performed as part of a job, the effective level of expertise required for a job increases, thereby lowering the number of potential applicants and raising the wage premium for such type of jobs. Note, however, that the level of expertise is not only a technological question but can depend on regulation such as occupational licensing. In this case, GenAI would impact the pool of experts only indirectly, for instance by lowering the cost of obtaining a license (e.g. by facilitating taking licensing exams).

There are some limitations to the task-approach. The framework's static nature prevents it from addressing the dynamics of labor market adjustments to technological shocks. By using comparative statics, the model compares pre-shock and post-shock steady states while overlooking the transition between them. Moreover, baseline task models (except \citealp{restrepo15}) assume frictionless, perfectly competitive labor markets with full employment. When AI displaces workers from tasks, the models assume they seamlessly transition to other work at potentially lower wages. In reality, displaced workers face unemployment spells due to search frictions, skill mismatches, or downward wage rigidity during potentially slow transitions.

These limitations can be addressed by embedding the task model in dynamic frameworks that incorporate occupational \cite{delRio2021,bocquet24} and sectoral transitions \citep{acemogluKongRestrepo24}. Displacement effects push workers to reallocate across labor markets, creating wage spillover effects that trigger further task reassignments between capital and labor. These dynamics can be formalized via a "propagation matrix" or network of skill similarity across occupations\cite{dabed2025equalising,Mealy2018}, capturing outflows from origin markets and inflows into destinations. Similarly, introducing multiple sectors linked by input-output structures \citep{jackson2019automation} captures how different sectors experience varying rates of technological change, leading to diverse productivity responses and differential wage growth that triggers sectoral reallocation. Since unemployment and transition costs represent welfare losses, accounting for these frictions could significantly alter the welfare assessment of technological shocks, with important implications for optimal policy \citep{Beraja24}.

While the framework's flexibility in modeling task reallocation is a strength, it also presents empirical challenges. Accurately identifying and measuring tasks, particularly as they evolve with technology, remains difficult. The distinction between routine and non-routine tasks that anchored earlier automation studies proves insufficient for understanding GenAI's impact, as AI can now automate previously non-routine cognitive tasks through pattern recognition and implicit rule learning \citep{ernst19}. This motivates the need for new taxonomies distinguishing simple from complex tasks to better characterize GenAI's labor market effects.

\subsection{Teams and work organisation\label{sec:collective_intelligence}}

The economics literature on automation and the impact on employment discussed in the previous sections has looked at jobs individually, considering how a new technology can affect jobs and tasks one-by-one. In recent years, a literature emerged out of a larger field of social choice theories combined with management theories that broadened the scope to understand how group performance can be enhanced by technology. Generically dubbed "Collective intelligence" (CI), it grew out of considerations around the wisdom of crowds, which had triggered interest already several centuries ago, famously discussed by Condercet and his jury theorem.\footnote{\url{https://en.wikipedia.org/wiki/Condorcet\%27s_jury_theorem}}

Wisdom of crowd effects rely on the statistical law of large numbers. It refers to the idea that large, diverse groups can make more accurate judgments than individuals, even if these individuals are experts \citep{surowiecki04}. A few conditions are required, however, for this effect to materialize. Most importantly each member of a crowd is giving an assessment independently from everybody else so that interaction between all individuals are minimal. In this case, a simple aggregation of all judgments will be more precise than any individual one. However, while individual errors can be eliminated by such an aggregation, systematic biases that might affect the crowd as a whole will not \citep{page17}. These effects have attracted significant interest during the previous wave of digitalisation as it became much easier to reach out to a larger number of people while designing protocols that would allow to better control for crowd selection and bias reduction. Consequently, the industry of opinion polls took off rapidly \citep{stephens18}.

However, in many real-world situations, independence between individual judgments is not guaranteed. Most obviously, when people work together in teams or within companies with regular exchange and communication, their interaction will be regular and warranted. An active field of research has developed around CI to measure and better understand, under which conditions, teams would operate more effectively and efficiently than individuals in isolation \citep{woolley10}. Importantly, issues such as the composition of teams, the individual characteristics of team members, the type of interaction they are entering and the type of tasks teams are supposed to execute play a key role in determining whether teams perform better together than their members separately \citep{malone18}. In particular, the complexity of tasks as measured by the number of components and the extent and structure of interdependencies between them significantly impacts the degree to which team performance exceeds that of individuals \citep{almaatouq21}. Moreover, for CI to be effective, group size might not be the most relevant factor; rather, the composition of the group and the way group members are being selected and interact with each other is important.\footnote{Much of the literature is empirical from the outset with very few generalizable results, in contrast to the economics of automation literature reviewed above. A few general considerations apply, however, and offer insights into how technology in general and AI in particular is likely to affect team production. In particular, teams benefit from internal diversity even if individual members are not top-performers in their respective field \citep{hong04,page08}. This "diversity bonus" stems from the fact that the accuracy of a prediction or execution of a task is the sum of individual minus the diversity of predictions/executions, linked to the K\"{o}nig-Huygens theorem in relation to the derivation of the variance of a random variable:
\[
\left(c-\theta\right)^{2}=\underbrace{\frac{1}{n}\sum_{i=1}^{n}\left(s_{i}-\theta\right)^{2}}_{\text{Average Error}}-\underbrace{\frac{1}{n}\sum_{i=1}^{n}\left(s_{i}-c\right)^{2}}_{\text{Diversity}}
\]
where $\theta$: the true value, $c$: the collective prediction, $s_i$: the individual predictions and $n$: the number of group members.}

Tasks are being differentiated between simple and complex ones, with some parallels to the expertise introduced by \cite{autor25}. For simple tasks for which it is easy to identify best performers (e.g., athletic champions), this diversity bonus might be small and the collective outcome worse than the best performer of the group, $\hat{s}_i$. When tasks become more complex, however, team performance almost always beats individual high-achiever \citep{almaatouq21}. A specific aspect of task complexity that influences this outcome is not only the number of components that need to be completed within each tasks but the way in which these components interact \citep{burton21,almaatouq21b}. More generally, the network structure of interactions -- whether at the task level or the level of individual members of a team -- play an important role in collective outcomes \citep{castro25}. For exmaple, bots can speed coordination or hinder it depending on network placement \cite{tsvetkova2024new}. Moreover, diversity is a multi-dimensional concept. Not all dimensions of diversity will be relevant for each task to be completed. \cite{cui24} distinguish between surface- and deep-level diversity. The first type encompasses observable categories such as age, gender and ethnicity whereas the second type includes differences in personality, values and cognitive styles. Depending on the tasks to be executed only a few dimensions might be relevant in generating a diversity bonus.

The use of digital technologies and AI in particular can impact this diversity bonus along three dimensions \citep{Piskorski2023,cui24,riedl25}: (i) it can add diversity, especially if certain dimensions are underrepresented; (ii) it can modify the structure of interaction between group members (adding or removing edges) that can improve the internal communication and information processing; and (iii) it can provide support services such as improve collective memory of essential information. Moreover, within a specific role, AI can aid in several functions such as improved memory, attention and reasoning \citep{Woolley2024}. Some experiences suggest that there are combinations of (i) and (ii) that are particularly effective to improve in-group communication. \cite{traeger20}, for instance, demonstrate how AI-agents with particular communicative styles can help intensify in-group communication. Similarly, as a teammate AI enhances memory, serving as a knowledge repository, which is especially valuable for diverse teams. In terms of attention, LLMs can help team dynamics by democratising speaking time and making sure all people are heard \citep{Reitz2024}. Finally, AI might contribute to reasoning and goal clarity within teams, which is considered the most challenging function, as it requires AI to understand the intentions, motivations, and beliefs of other team members—what is often referred to as a "theory of mind" \citep{Woolley2024}. 


Ultimately, the extent to which generative AI influences teams will depend on its adoption. For AI to be effective, team members must trust and collaborate with it. However, there is a delicate balance: enough trust is necessary for AI to integrate into the team, but over-reliance on AI could undermine human judgment and collective intelligence \citep{Woolley2024}. Techniques such as cognitive forcing functions \citep{Bucinca2021} and providing explanations \citep{Vasconcelos2023} for AI outputs can help achieve this balance, promoting thoughtful use while preserving collective intelligence.

\section{Exposure of jobs to AI: Ex-ante analyses\label{sec:AI-jobs}}


Most available evidence focuses on ex-ante effects of the introduction of AI on jobs, so-called "exposure measures". In this sub-section, we start by unpacking these AI exposure measures, detailing how they are constructed and what they aim to capture (subsection \ref{subsubsec:identify}-\ref{subsubsec:GenAI}). Secondly, we evaluate the strengths and limitations of each measure, providing guidance on their appropriate use (subsection \ref{subsubsec:StrengthsWeaknesses}). Finally, we review their general predictions on which groups of workers are likely to be most impacted by AI technologies. Moreover, we compare different measures of occupational exposure to automation (subsection \ref{subsubsec:Comparison}). For readers primarily interested in these predictions and less concerned with the technical details, we recommend skipping the first subsections and proceeding directly to the last one.

\subsection{Identifying task-level information\label{subsubsec:identify}}

Building on the task framework, most measures of exposure to AI focus on tasks. These measures assess whether specific tasks—such as writing a literature review—can be automated by AI algorithms. While evaluations start at the task level, they can be aggregated to the occupation level, providing exposure metrics for various population groups, categorised by wage group, gender, age, and ethnicity. This approach can help predict how new technologies can alter the task composition within occupations.

Information about different tasks is available only for a few countries on a regular basis: e.g. the United States, Italy, Germany. Most studies collect task-level information from the Occupational Information Network (O*NET) database, a comprehensive resource developed by the U.S. Department of Labour.\footnote{\url{https://www.onetonline.org}} O*NET provides detailed information on the tasks, skills, abilities, work contexts, and knowledge required across occupations in the U.S. labour market. Measures of AI exposure differ depending on which O*NET variables are used—for example, \cite{Pizzinelli2023} focus on work environments, while \cite{Autor2024a} use job titles.

O*NET is particularly valuable because it offers detailed task-level data and task importance weights, enabling researchers to aggregate task information into occupation-level metrics. Starting from "detailed work activities" like "analysing data" or "writing reports," these can be aggregated into broader "tasks" and then to occupations classified at the 6-digit SOC (Standard Occupational Classification) level. For example, the skill "complex problem solving" might have a high importance score for software engineers, indicating its relevance within that occupation.

Few other countries have databases as detailed as O*NET. Researchers studying the impact of AI on occupations outside the U.S. often use crosswalks to link U.S. data with international classifications such as ISCO (International Standard Classification of Occupations). However, this approach presents challenges due to differences in occupational structures across countries, especially between high- and low-income nations. For example, the role of an IT developer in the United States might differ substantially from that in India, leading to variations in AI exposure measures, as highlighted by \cite{Carbonero2023}.

An alternative approach, suitable for a selected number of developing countries is to use skill surveys, such as the World Bank STEP survey, which identifies a series of competences required for different jobs. For instance, \cite{Carbonero2023} use natural language processing (NLP) techniques with the help of BERT and SBERT to build crosswalks between O*NET -- where AI exposure measures already exist -- and other occupational classifications in both developed and developing countries. They compute the semantic similarity between tasks described in O*NET and skill descriptions from surveys like STEP (for developing countries) and PIAAC (for developed countries). This approach captures the overlap between tasks performed by U.S. workers and those in other countries, allowing for a broader, more global application of AI exposure measures. Similarly, the PIAAC survey provides an overview of competences among adult workers in OECD countries that can be used to complement skill cross-walks for advanced economies. However, even when combined STEP and PIAAC surveys together allow to estimate exposure rates only for 53 countries, with potentially outdated information on task compositions as the STEP surveys have been administered at the beginning of the 2010s. \cite{lewandowski25}, therefore, suggest a regression-based approach to extend occupational profiles by considering country-endowments to expand the country coverage to 103 countries.

\subsection{Leveraging expert insights\label{subsubsec:experts}}

Identifying tasks performed by different occupations is only a first step. To assess the exposure of tasks to advances in AI, a first strand of the literature directly leverages insights from experts. By directly consulting individuals with specialized knowledge, researchers aim to gauge which tasks or occupations are susceptible to automation. The approaches differ in terms of who they ask, what they ask, the level of aggregation, and whether they employ an expert-only or hybrid method.

\cite{Frey2017} employed a hybrid method that combined insights from AI experts with the predictive capabilities of machine learning to predict which occupations were susceptible to automation. Specifically, they consulted a group of machine learning and AI experts to evaluate the automatability of occupations. Experts were asked to hand-label a selection of occupations as automatable or not, based on whether the tasks involved could be performed by existing or foreseeable technologies in the near to medium term (generally within a couple of decades). Importantly, their assessment was at the occupation level, considering whether whole occupations could be automated, rather than individual tasks. A machine learning classifier was then trained using this data, employing features from O*NET that describe the tasks and skills of occupations. They identified key "bottlenecks" to automation, such as tasks requiring perception and manipulation, creative intelligence, and social intelligence. While this approach is innovative, it has limitations—primarily, the classifier's accuracy is unclear, as we lack information on whether it correctly categorized occupations as automatable or non-automatable. More detailed expert insights would be needed to verify the machine learning model's classifications.

\cite{Brynjolfsson2018} used a crowd-sourcing approach, gathering insights from workers in affected sectors. These participants were familiar with specific job tasks, providing a practical perspective on the feasibility of automation in their respective fields. They developed a measure called "Suitability for Machine Learning" (SML). Participants were asked to evaluate to what extent individual tasks within their occupations could be performed by machine learning (ML) technologies. The focus was on the feasibility of automating tasks using current or near-future ML capabilities, without specifying a precise time horizon. Task evaluations were then aggregated to the occupation level using task importance weights from the O*NET database, allowing for an analysis of occupations based on the composition of their tasks. One challenge is that workers may either overestimate or underestimate the automation potential due to fear of job loss or lack of awareness of AI capabilities. Additionally, reliance on non-expert opinions about ML technology might affect the accuracy of assessments.

\cite{Felten2018, Felten2021, Felten2023a} combined insights from AI experts with occupational data to assess the exposure of occupations to AI. First, they asked AI experts from Electronic Frontier Foundation (EFF) to evaluate the progress of AI capabilities in specific application areas, such as image recognition or language modelling.  Second, the authors assessed how these AI capabilities related to the abilities required in different occupations by using a crowd-sourced matrix of relatedness collected via Amazon Mechanical Turk. Specifically, the authors mapped AI capabilities to occupational abilities defined in O*NET—such as oral comprehension, oral expression, inductive reasoning, and arm-hand steadiness. Finally, these abilities were aggregated to the occupation level using O*NET importance weights. The resulting measure, known as AI Occupational Exposure (AIOE), quantifies the overlap between AI applications and the abilities required for various occupations. 

However, this approach has potential limitations. On the one hand, experts may be overly optimistic about AI progress, introducing an optimism bias into their assessments. On the other hand, by focusing solely on current AI capabilities, the measure may underestimate future advancements, failing to capture the full potential impact of AI on occupations. Additionally, the crowd-sourced mapping between AI capabilities and occupational abilities might not capture the full complexity of tasks, as it relies on general perceptions that may oversimplify nuanced job requirements and work processes in their entirety. 

The binary classification (automatable or not) probably oversimplifies the spectrum of automation potential. Moreover, assessing exposure at the occupation level may ignore the variability of tasks within occupations - a key adjustment margin according to the task framework. The bottlenecks to automation – creative skills – might also not be so relevant anymore, with gen-AI making huge progress in image- and language-generation tasks. Finally, the studies’ predictions are based on the technological feasibility of automation, not accounting for economic feasibility, regulatory constraints, or social acceptance. 

\subsection{Leveraging patent data\label{subsubsec:patents}}

Another strand of the literature uses patent data to assess how AI could impact the labour market. By analyzing the textual content of patents and comparing it to occupational tasks, researchers can estimate which jobs are more susceptible to automation. Compared to expert-based measures, patent-based approaches have two advantages: (1) they focus on AI applications that are already developed and potentially adopted, and (2) they are not subject to the biases inherent in expert opinions.

The key studies in this area differ in both their selection of patents and the text analysis methods they use to quantify semantic similarity. Some researchers focus on specific subsets of patents, such as those classified as disruptive or explicitly related to AI technologies. Additionally, the techniques for measuring similarity between patent texts and occupational tasks vary—from basic keyword matching to sophisticated natural language processing algorithms. 

\cite{mann23} were among the first to use U.S. patent data to study the impact of automation on the labour market. They classified patents from 1976 to 2014 as either automation-related or not. Building on \cite{Frey2017}, they manually labeled hundreds of patents and then trained a machine learning algorithm to predict whether remaining patents were related to automation, using natural language processing techniques. However, like Frey and Osborne’s approach, the classification process is somewhat opaque, and it is unclear how accurate the algorithm’s predictions are. Additionally, their work focuses on patents before 2014, preceding the recent rise of AI. Unlike much of the literature, they link patents to industries rather than occupations, providing a broader view of automation's sectoral impact.

\cite{Dechezleprêtre20} and \cite{Gathmann2022} use European patent data to create industry-level measures of exposure to automation technologies, but they apply different methods of identifying automation-related patents. \cite{Dechezleprêtre20} classify patents from 1997 to 2011 as automation-related based on the frequency of relevant keywords exceeding a certain threshold. In contrast, \cite{Gathmann2022} use a combination of patent codes and keyword searches to classify European patents over a broader period from 1990 to 2018. These methods are more transparent than the approach used by \cite{mann23}, as they rely on machine learning algorithms to classify patents. Like \cite{mann23}, both studies focus on earlier periods when AI technologies were still in their early stages, providing valuable insights into the initial phases of automation. 

Another pioneering work in this field is \cite{Webb2020}, who assesses the potential impact of AI on occupations by leveraging patent data and occupational task descriptions. He begins by identifying AI-related patents through a search for specific keywords and classifications indicative of AI technologies, such as "neural network" or "unsupervised learning." From the titles of these patents, Webb extracts verb-noun pairs that represent actions (verbs) and objects (nouns) associated with AI advancements—for example, (diagnose, disease). He then compares these verb-noun pairs to those found in the task descriptions of occupations from the O*NET database. Tasks exhibiting a high overlap in verb-noun pairs are considered more exposed to AI. This task-level exposure is subsequently aggregated to the occupation level using O*NET importance weights, enabling an evaluation of which occupations are most susceptible to automation by AI.

Webb's methodology offers the advantage of direct matching between technological capabilities and occupational tasks, capturing granular details of how AI might affect specific job components. Compared to other natural language processing methods, this approach is much less of a "black box," providing greater transparency in how exposure is determined. However, relying solely on verb-noun pairs might miss nuanced semantic relationships and contextual meanings within the texts. Additionally, identifying AI patents based solely on keywords may not capture all relevant technologies or might include patents not directly related to AI.

To refine the selection of AI-related patents, \cite{kogan23} examine the impact of technological innovation on labour markets by focusing on a specific subset of patents known as "breakthrough" patents. They identify these patents using the methodology of Kelly et al. (2021), who define breakthrough patents as being both (1) novel (not cited before) and (2) impactful (cited frequently afterwards). The authors link these disruptive patents to industry-occupation cells. However, concentrating solely on highly cited patents might miss emerging technologies that have not yet accumulated citations but could have substantial future impacts on the labour market.

Another strand of the literature leverages recent advances in textual analysis to improve the quantification of patent-task overlap. \cite{Prytkova2024} propose a method to assess the impact of AI on occupations using advanced natural language processing techniques. They use sentence transformers, a type of deep learning model that generates contextualized embeddings for sentences or paragraphs, capturing nuanced semantic meanings. The authors process both the textual content of AI-related patents and the task descriptions from the O*NET database to create these embeddings. This approach allows for a more accurate and context-aware matching compared to simpler text analysis methods, as it considers the context and relationships between words within sentences.

Similarly, \cite{Septiandri2024} present a methodology for assessing the risk of job automation by leveraging advanced natural language processing models. They employ Bidirectional Encoder Representations from Transformers (BERT) to generate embeddings for both patent documents and occupational task descriptions. BERT is a state-of-the-art deep learning model that captures the context of words from both directions in a sentence, enhancing the understanding of semantic relationships.

These approaches offer improved accuracy over traditional methods by effectively capturing the contextual meaning of text and the nuanced relationships between technological capabilities and job tasks. However, the complexity of neural networks can make it challenging to interpret the exact reasons behind specific similarity scores, potentially reducing transparency.

\subsection{Leveraging GenAI\label{subsubsec:GenAI}}

More recently, another strand of studies has used gen-AI itself to assess the impact of AI on workers. Compared to the other methods, this approach presents two main advantages: (1) it is very easy to use and implement, and (2) it leverages a very wide range of information to determine exposure, the model being trained on a very vast sample of texts. 

\cite{Eloundou2023} were one of the first papers to directly use gen-AI models to measure worker exposure to gen-AI. Specifically, they asked ChatGPT to assess whether specific tasks could be automated by generative AI. Each task from O*NET was presented to ChatGPT, which was asked to determine if it could perform the task with at least the same quality and in 50\% less time than a human worker. Tasks meeting this criterion were labeled as having high exposure to automation. The exposures at the task level were then aggregated to the occupation level using O*NET importance weights, allowing the researchers to calculate an exposure score for each occupation.

\cite{Gmyrek2023, Gmyrek2025} extended Eloundou et al.’s methodology to a broader range of countries by utilizing information available in the International Standard Classification of Occupations (ISCO). ISCO is an occupational nomenclature developed by the International Labour Organization (ILO) and used in 73 countries, making it useful to compare exposures to AI across national contexts. The authors asked Chat-GPT to assess the automation risk from gen-AI for all tasks from the ISCO nomenclature. Interestingly, they requested ChatGPT to justify its decisions, thereby providing some transparency into the model’s decision-making process. However, one limitation of ISCO compared to the US O*NET database is that it does not provide weights to aggregate tasks back into occupations. To address this, the authors weighted each task within an occupation equally. 

In another study, \cite{kogan23} also used GenAI models to assess the impact of AI on workers. Adopting an approach similar to \cite{Eloundou2023} they asked ChatGPT to evaluate whether specific tasks could be performed by AI technologies either autonomously (substitution measure) or with human assistance (complementarity measure). They considered both current capabilities and anticipated future advancements. The task-level assessments were then aggregated to the occupation level using importance weights, resulting in exposure scores for each occupation.

Finally, similar to \cite{Eloundou2023} \cite{chen25} make use of the occupation descriptions in the Chinese Occupational Classification Dictionary for China in 2022. They employ three different large language models (Open AI's GPT-4, InternLM by Shanghau AI Laboratory and Sense Time, and GMT) to select occupations based on their description into four different categories depending on their exposure level and the feasibility of automating the underlying occupation. Given the lack of detailed task descriptions, \cite{chen25} apply their analysis directly at the occupational level, which limits the possibility for a more granular approach. Their results point to non-routine cognitive occupations as the most exposed whereas non-routine manual occupations and cognitive interpersonal occupations receive the lowest exposure scores. 

Relying on AI models like ChatGPT to evaluate their capabilities may introduce self-assessment biases, as the models might overestimate or underestimate their actual performance. Additionally, the assessments are not validated by human experts. Rather than introducing new insights, such AI-generated assessments are more likely to reflect aggregate patterns or dominant narratives derived from their training data — which consists of a vast range of online text — rather than a grounded evaluation of work processes and human roles. Moreover, with the exception of \cite{chen25} none of the authors have executed robustness checks by using different Gen-AI models or different vintages from the same model to check whether results are consistent. The robustness check carried out by \cite{chen25}, in particular, suggests the need for caution when interpreting the findings and highlights the importance of external validation to ensure accurate assessments of AI's impact on the labour market.

\subsection{Strengths and limitations of exposure measures\label{subsubsec:StrengthsWeaknesses}}

The wide variety of approaches to measuring the AI exposure of occupations makes it challenging for researchers and policymakers to decide which method to adopt. Nevertheless, understanding the nuances of these different approaches is essential for accurately interpreting their results and their implications for the labour market.

Each of the individual  methods presented in sections \ref{subsubsec:identify}-\ref{subsubsec:GenAI} has its advantages and limitations, which we discuss in more detail in the following. Before we elaborate on the strengths and limitations of each individual AI exposure measure, we highlight five main limitations for economic policy that concern all AI exposure measures and which we already hinted at in the previous discussion of the task-approach: 

Firstly, the exposure measures are based on the tasks of an existing, unaltered work process. Highly impactful technologies are often intended to change the work process itself and major organizational changes are necessary to use the technology effectively and efficiently. Hence, the AI exposure measures are built on a set of tasks that the AI is likely to destroy or to fundamentally alter. A high or low exposure measure on a specific task that does not exist anymore because it has become irrelevant (i.e., not replaced by AI) in the new work process does not say anything about the risk of the worker linked to that task after the adoption of AI. Exposure scores that reflect such risks would have to anticipate how AI modifies the tasks of a whole work process in a specific setting, which they do not do.

Secondly, the derivation of AI exposure measures is not linked to any macroeconomic variables, such as wages, profits, investment costs, adjustments costs, prices, or productivity. Rather, the exposure measures consider a fixed-coefficient (Leontieff) technical relationship between tasks and technology. The macroeconomic implications of determining the technological feasibility of AI carrying out certain human tasks therefore remain unclear and vague. Introducing estimates for cost curves indicating when AI will under-cut a range of human competencies, \cite{peng25} demonstrate important cross-sectoral differences in both the extent to and the time-frame by which a significant portion of jobs can be substituted with machines.

Thirdly, and related to the previous point, the AI exposure measures do not indicate any likelihood of adoption of the technology. For a variety of purely economic factors (e.g., relative comparative advantages between human and AI) but also for socioeconomic, legal, ethical or cultural reasons, AI technology might or might now be adopted for certain existing tasks or differently across countries. The AI exposure measures do not provide any indication about these considerations.

Fourthly, the AI exposure measures remain subjective, expert-based (or sometimes AI-based) opinions of what AI or GenAI can probably do today in terms of carrying out an abstract activity, or what it could probably do in the near or midterm future. As AI evolves and as experts adapt their views, exposure measures may change or turn out to be biased or simply wrong. 

Finally, and this point can also be seen as a consequence of the previous four points: There exists no commonly accepted interpretation of what the AI exposure measures actually mean. Some authors are more explicit and call them “risk of automation”, or “suitability to machine learning”, or make a distinction of “automation” and “augmentation” but ultimately, it remains unclear whether a high AI exposure means that the occupation will disappear, or change, whether a person becomes unemployed, or needs to re-train to perform on the same job, or re-train for an entirely different job, and whether higher or lower skill levels are need. These interpretations, which are extremely relevant for policy makers, are up for discussion.

\paragraph{Old vs. new generations of AI}

Early measures of AI exposure, such as those by \cite{Brynjolfsson2018}, \cite{Frey2017}, and \cite{Felten2018}, focus on AI applications in pattern recognition and predictive tasks. These technologies were primarily specialised in automating routine or repetitive tasks. However, recent advances in AI, notably with generative AI, have expanded AI's capabilities, enabling the automation of more complex tasks that require creativity or social intelligence.

For example, \cite{Frey2017}, \cite{Felten2018,Felten2021} have identified "bottlenecks" in AI substitution, notably creative skills and social intelligence, which are considered difficult to automate. However, the latest generations of generative AI have already begun overcoming these obstacles, challenging the conclusions of earlier measures. Thus, more recent measures are better equipped to capture the effects of the latest AI advancements, reflecting the changing reality of automation potential. This evolution highlights the importance of continuously updating AI exposure measures to account for rapid technological progress. Methods based on current AI models, such as those by \cite{Eloundou2023}, offer a more accurate view of the current AI's potential impact on various professions.

\paragraph{Prospective vs. Retrospective Measures}

Methods also differ in their temporal horizon. Some measures evaluate the feasibility of future automation, without guaranteeing that the technology will actually be adopted. For example, \cite{Brynjolfsson2018} ask experts to assess the impact of current or near-future AI capabilities on workers, while \cite{Frey2017} assess the automation risk over the next couple of decades. However, factors such as high implementation costs or low output quality can hinder adoption, meaning these measures capture only the potential for adoption. 

At the other end of the spectrum, some measures are based on historical trends in AI advancements, thus sidestepping the question of whether the technology will be eventually adopted. For instance, \cite{Felten2021,Felten2023a} or patent-based methods (e.g., \citealp{Webb2020,kogan23}), focus on technologies that are already present and being innovated upon. However, these methods may only be a lower-bound estimate of AI exposure, and not fully reflect the disruptive potential of future technological developments.

\paragraph{Substitution vs. Complementarity}

Another important distinction among the measures is their focus on the substitution or complementarity effects of AI. Some studies, like \cite{Frey2017}, \cite{Brynjolfsson2018}, and \cite{Eloundou2023}, specifically concentrate on AI's potential to replace human labour, thus assessing the risks of job automation.

Other research adopts a more nuanced approach, recognizing that AI can also have complementary effects by enhancing workers' capabilities rather than replacing them. For instance, \cite{Felten2018,Felten2021,Felten2023a}, \cite{Webb2020} and \cite{Prytkova2024} remain agnostic regarding substitution, leaving the door open for augmentation effects. \cite{Fossen2019} interpret Frey and Osborne's measure as an indicator of automation potential, while they consider Felten et al.'s AIOE (AI Occupational Exposure) as a measure of complementarity potential.

An emerging strand of literature aims to focus directly on AI's augmentation effects. \cite{Brynjolfsson2018} identify complementary occupations as those with high variance in task exposure (i.e., a combination of tasks with both high and low exposure within the occupation). \cite{kogan23} directly query AI models to determine which tasks would require human assistance. \cite{Pizzinelli2023} measure complementarity effects using work environment data from O*NET, defining complementary tasks as those with high stakes and a low probability of being left unsupervised by humans (such as flying an aeroplane). Finally, \cite{Autor2024b} measure complementarity by analysing the textual overlap between patents and occupational outputs, rather than tasks.

This distinction between substitution and complementarity is crucial for understanding AI's real impact on employment. While some tasks may be fully automated, others can benefit from AI assistance, improving productivity and creating new opportunities for workers.

\paragraph{Expert-Based vs. Machine Learning-Based Measures}

Measures based on human evaluations, whether from AI experts or through crowdsourcing, offer transparency but may be subject to sociological or professional biases. For example, AI experts might have an imperfect understanding of tasks performed in other professions, which could influence their assessments.

In contrast, machine learning-based measures are potentially less susceptible to certain human biases, although algorithmic biases are a real concern too. These methods are often less transparent in how they identify exposure: does the machine capture a statistical artefact or true exposure? Additionally, some predictions can be surprising, such as the idea that cooks and waiters have a high likelihood of being automated according to Frey and Osborne's measure.

Therefore, it's important to consider the strengths and limitations of each approach. Expert-based methods can benefit from human experience and judgment while machine-learning approaches can process large amounts of data and identify complex patterns. A combination of both could offer a more balanced assessment of AI exposure.

\paragraph{Developed vs. Developing Countries}

Most AI exposure measures are estimated using U.S. data on skills, which can pose problems when applying them to other countries. The task content within professions can vary significantly between countries, especially between developed and developing economies. Therefore, directly using these measures in developing countries may not accurately capture the actual exposure of occupations to AI.

There is a research gap in looking at the effect of AI in developing countries. The notable exception is \cite{Gmyrek2023,Gmyrek2025}, who leverage AI to assess which jobs can be replaceable or complemented with AI.\footnote{The original study refers to substitutable jobs as automatable and complementary as augmented.} This research finds that while lower-income countries may have less substitutable jobs, they also have less complementary jobs. Using employment data from 59 countries, they find that in low-income countries (LICs), only 0.4\% of jobs fall into the category of high substitution potential, compared to 5.5\% in high-income countries (HICs). Conversely, lower-middle-income countries (LMICs) have the highest share of jobs with complementary potential, at 14.4\% of total employment. However, these potential gains from complementarity may not be realised by infrastructure challenges, such as limited access to digital tools and skills. The study also highlights that in HICs, women are disproportionately affected by automation, with 8.5\% of female employment classified as high automation potential, compared to 3.9\% for men. Similarly, \cite{lewandowski25} expands AI exposure scores by using a regression approach to estimate occupational profiles for around 103 countries. While not distinguishing between substitution or augmentation, their results confirm other estimates about the larger share of occupations in advanced economies to be exposed to AI in comparison to low- and middle-income countries. These findings underscore the importance of considering national contexts—such as education, technological infrastructure, and professional practices—when evaluating AI’s potential impact on labour markets. 

\subsection{Which types of workers are most exposed to AI?\label{subsubsec:Comparison}}


AI and more recently GenAI have prompted concerns about job displacement and income insecurity. What is new with AI is that while previous waves of technological changes, such as robots and software technologies, affected manual and abstract routine jobs respectively, AI has started affecting non-routine occupations which once were considered not automatable \citep{Autor2024b}. However, David Autor contends that expert commentators and journalists tend to overstate the extent of job substitution and ignore the strong complementarities between automation and human labour that increase productivity, raise earnings, and augment demand for labour. 

Recent studies have shed light on which occupations are most exposed to the advancements in AI. Using the literature on AI exposure measures that we discussed in the previous sections, including its limitations, we provide an overview of which types workers or which occupations could be mostly exposed to AI. In particular, exposure varies not only depending on the type of occupation but also regarding the demographic and socio-economic characteristics of those occupying these jobs.

\cite{Eloundou2023} find that occupations requiring programming and writing skills, such as consultants and software developers, are at a high risk of automation. These roles often involve tasks that can be performed by generative AI models, making them susceptible to technological substitution. In contrast, occupations that require analytical and scientific skills appear to be less exposed to AI for now. These roles often involve complex problem-solving and critical thinking tasks that are currently harder for AI to replicate.

Location is another factor explaining AI exposure. In this respect, workers in urban areas and those predominantly working from home are more exposed to GenAI because cities are hubs for innovation and individuals who primarily work from home may use technology more intensively \citep{Nurski2024}.

Gender plays a key role in determining exposure levels. Women are likely to be more affected \citep{Webb2020,genz23,Pizzinelli2023}. \cite{Cazzaniga2024} find that women are generally more exposed to GenAI. Other studies make a distinction: women are exposed to GenAI language modelling, followed by AI in general and less to image generation GenAI. Men exhibit the opposite trend \citep{Nurski2024}. This raises concerns about the potential gender disparities resulting from AI adoption.

Exposure changes also over the life cycle. Younger prime-age workers (age 25-44) have a higher likelihood of exposure \citep{Nurski2024}. Older workers are potentially less able to adapt to the new technology \citep{Cazzaniga2024}, especially in emerging economies where older workers have benefited from less years of schooling than their younger colleagues. Most recent evidence suggests that in certain professions such as software development and paralegal entry-level jobs are being replaced at an increasing rate.\footnote{\url{https://www.nytimes.com/2025/05/30/technology/ai-jobs-college-graduates.html}, \url{https://medium.com/design-bootcamp/the-end-of-prestige-how-ai-is-quietly-dismantling-the-elite-professions-0b96b649edf4}}

Highly educated and high-skilled workers are mostly exposed \cite{Webb2020,Eloundou2023,Pizzinelli2023,Cazzaniga2024}. However, there might be different implications in advanced and emerging economies, since high-skill service occupations constitute a large chunk of employment in advanced economies, but much less in developing countries. 

This has implications for wages. \cite{Webb2020} and \cite{Eloundou2023} find a non-monotonic relation between wage and exposure: it is first increasing in wage, but then slightly decreasing for very high-wage occupations. By contrast, older papers \citep{Brynjolfsson2023b, Frey2017} find a negative correlation with wages. \cite{Fossen2022} distinguish between investment in software, robots and AI and find a positive correlation between AI and wages for the United States, using Brynjolfsson's SML scores as AI exposure measure, and a negative correlation for  the other two types of innovation. Some recent studies point out that complementarities between AI and high-income workers can play a key role in shaping the relationship between wages and exposure and on inequality. Workers in the upper tail of the earnings distribution are more likely to be in occupations with high exposure, but there is also high potential complementarity \citep{Pizzinelli2023}.  If there is a strong complementarity between AI and high-income workers, labour income inequality may increase \citep{Cazzaniga2024}.

The labour force in advanced economies appears to be more exposed to displacement risks of AI and  GenAI than in emerging economies, but at the same time is also more prepared to exploit benefits \citep{Gmyrek2025,lewandowski25}. Emerging markets and low-income countries are less prepared to leverage AI, which could exacerbate cross-country income disparities \citep{Cazzaniga2024}.

\subsubsection{Quantitative analysis and comparison of AI exposure measures.\label{comparison_gross}}

The previous review and analysis of the literature has revealed that a variety of AI exposure measures exists. Besides general limitations of using AI exposure measures as predictors of specific economic consequences, as we previously discussed, the question arises whether the exposure measures point in the same direction regarding the extent to which occupations are exposed or if we see very different measures for similar jobs and occupations, depending on the exact methodology used. Therefore, we compare recent measures of occupational exposure to automation, arising from AI and GenAI.  Five distinct exposure measures are analysed. To ensure comparability, all measures are cross-walked and standardised to the 2018 Standard Occupation Classification (SOC).\footnote{By cross-walking all measures to the 2018 SOC, certain occupational properties may change. For instance, in \cite{Eloundou2023}, the O*NET 2019 codes 27-3043.05 (Poets) and 27-3043 (Writers) have distinct exposure scores. However, in the 2018 SOC, these occupations are merged, resulting in a slight reduction in the overall exposure score for writers.} We also incorporate mean and median wages provided by the Bureau of Labour Statistics (BLS), consistent with the data used in the replication materials of \cite{Eloundou2023}. Additionally, we create a global exposure index by normalising each measure from 0 to 1 and calculating their average. We refer to this as the ``Mean Normalised Measure".

Table \ref{tab:corrW_AI} shows the correlation between the six-digit SOC occupations and the log of median wages. Earlier automation exposure measures, such as those by \cite{Frey2017}, and \cite{Brynjolfsson2018}, show a negative correlation with wages. The correlation for Brynjolfsson et al. is notably weaker though still statistically significant. In contrast, more recent measures, including those by \cite{Webb2020}, \cite{Felten2021}, and \cite{Eloundou2023}, focus specifically on AI-related impacts and exhibit a positive correlation with wages, ranging from 0.27 to 0.54. Notably, \cite{Felten2021}'s measure shows the strongest positive correlation (0.54). The Mean Normalised Measure also shows a positive correlation with wages, although at a more moderate level.

\begin{table}[htb]
    \centering
    \caption{Correlation between wages and AI-exposure measures\label{tab:corrW_AI}}
    \bigskip
    \begin{tabular}{l|c} \hline \hline
       \textbf{Measure}  & \textbf{Correlation} \\ \hline
       \cite{Frey2017}  & $-0.5576^{***}$ \\
       \cite{Brynjolfsson2018} & $-0.0741^{**}$ \\
       \cite{Webb2020} & $0.2704^{***}$ \\
       \cite{Felten2018} & $0.5375^{***}$ \\
       \cite{Eloundou2023} & $0.4243^{***}$ \\ 
       \cite{Septiandri2024} & $0.1787^{***}$ \\
        \cite{Gmyrek2023} & $0.2811^{***}$ \\ \hline
        \textbf{Mean normalised measures} & $\mathbf{0.3043^{***}}$ \\ \hline \hline
       \multicolumn{2}{c}{\pbox[c]{10cm}{\medskip \footnotesize Note: Correlation between log median wages and exposure measures at SOC 2018 level; $^{***}: p < 0.001, ^{**}: p < 0.01, ^{*}: p < 0.05.$}}
    \end{tabular}
\end{table}

Table \ref{tab:top5occ} lists the top five occupations most affected by each exposure measure. There is substantial variation across measures in terms of the types of occupations identified. The exposure measures in \cite{Frey2017}, \cite{Brynjolfsson2018} and \cite{Webb2020} identify occupations requiring engineering skills or involving low-skill manual work as the most exposed to automation. In contrast, the measures presented by \cite{Felten2021} and \cite{Eloundou2023} are more aligned with cognitive and analytical tasks, with measures in \cite{Felten2018} linked to occupations involving mathematical analysis. The Mean Normalised Measure captures a mix, incorporating both cognitively demanding and low-skill occupations.

\begin{table}[htb]
    \centering
    \caption{Top five most exposed occupations\label{tab:top5occ}}
    \scalebox{0.5}{
    \begin{tabular}{p{2.8cm}|p{2.8cm}|p{2.8cm}|p{2.8cm}|p{2.8cm}|p{2.8cm}|p{2.8cm}|p{2.8cm}} \hline \hline
      Frey \& Osborne & Brynjolfsson et al. 2018 SML & Webb AI & Felten et al. AI & Elondou et al. GPT & Septiandri et al. AII & Gmyrek et al. Gen AI & Mean normalized exposure measures \\ \hline
      Data Entry Keyers (0.99) & Concierges (3.90) & Marine Engineers and Naval Architects (100.00) & Genetic Counselors (1.53) & Survey Researchers (0.84) & Cardiovascular Technologists and Technicians (0.64) & Data Entry Keyers (0.70) & Data Entry Keyers (0.76) \\
      Insurance Underwriters (0.99) & Mechanical Drafters (3.90) & Water and Wastewater Treatment Plant and System Operators (100.00) & Financial Examiners (1.53) & Interpreters and Translators (0.82) & Sound Engineering Technicians (0.57) & Word Processors and Typists (0.65) & Atmospheric and Space Scientists (0.75) \\
      Library Technicians (0.99) & Brokerage Clerks (3.78) & Astronomers (100.00) & Actuaries (1.52) & Public Relations Specialists (0.81) & Nuclear Medicine Technologists (0.53) & Billing and Posting Clerks (0.64) & Credit Analysts (0.75) \\
      Tax Preparers (0.99) & Gambling Cage Workers (3.77) & Physicists (100.00) & Budget Analysts (1.50) & Writers and Authors (0.79) & Air Traffic Controllers (0.52) & New Accounts Clerks (0.64) & Computer Programmers (0.73) \\
      Telemarketers (0.99) & Office Machine Operators, Except Computer (3.74) & Atmospheric and Space Scientists (100.00) & Procurement Clerks (1.49) & Animal Scientists (0.78) & Magnetic Resonance Imaging Technologists (0.52) & Loan Interviewers and Clerks (0.64) & Dispatchers, Except Police, Fire, and Ambulance (0.73) \\ \hline \hline
    \multicolumn{8}{c}{\pbox{10cm}{\medskip \footnotesize Numbers in parentheses correspond to the exposure measure. }}
    \end{tabular}}
\end{table}

Exposure does not necessarily mean that jobs are being replaced. For instance, \cite{baek25} find that the AI-exposure measures developed by \cite{Felten2018} are positively correlated with employment in the case of the Republic of Korea, one of the most highly robotised industrial economies \citep{carbonero18}. However, when controlling for other dependent variables, these correlations lose their statistical significance.

\begin{figure}
    \centering
    \caption{AI-exposure measures\label{fig:AIexposureMeasures}}
    \includegraphics[width=1.0\linewidth]{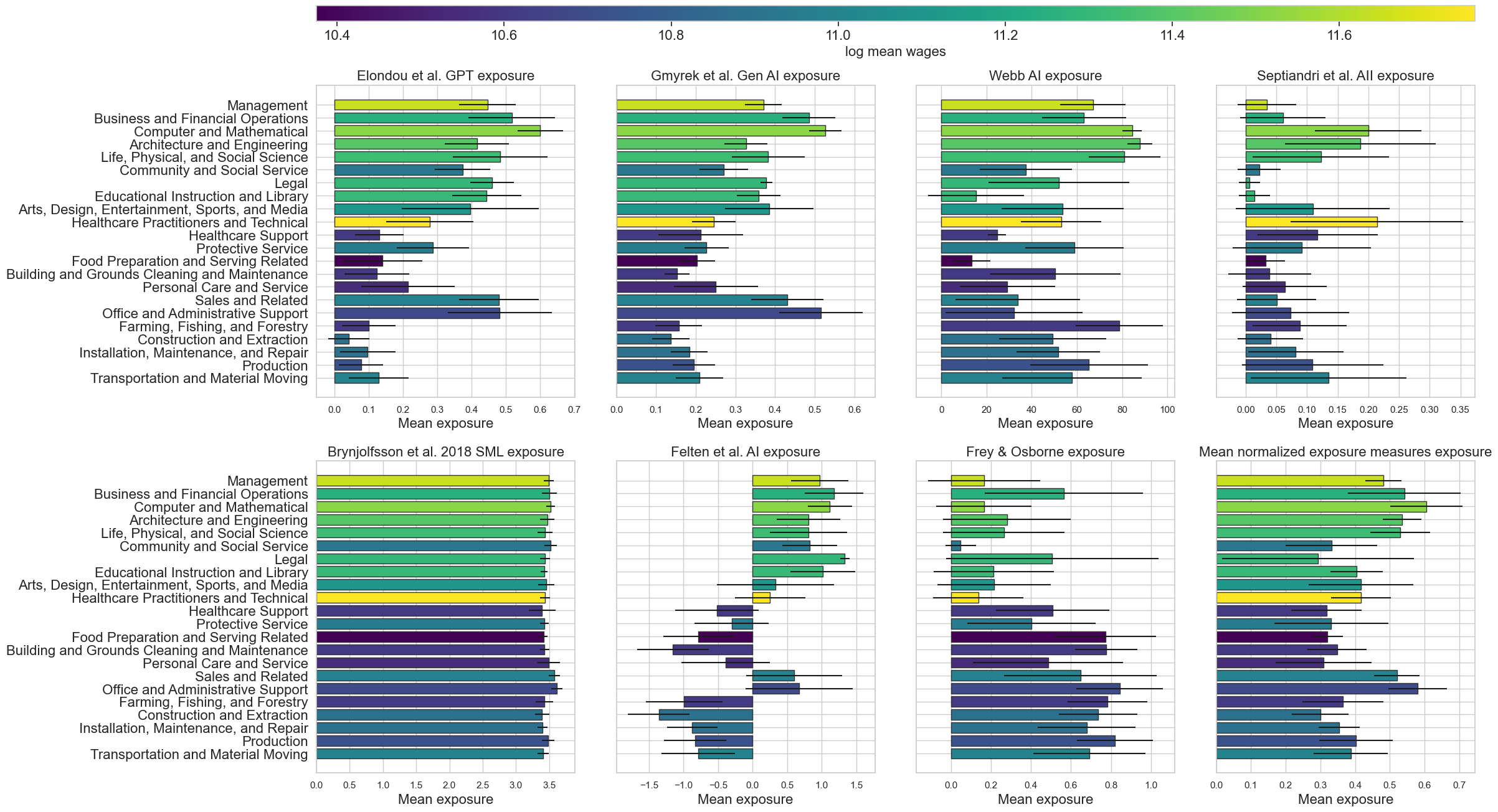}
    \pbox{15cm}{\footnotesize The chart compares mean AI-exposure measures at the 1-digit SOC 2018 category level across different studies. Each bar represents the mean exposure of the occupational category, averaged across all occupations within that category. Error bars indicate the standard deviation around this average. The colour represents the log mean wages of the occupational category, with lighter colours indicating higher wages.}
\end{figure}

To better understand how these exposure measures reflect skills and sectoral differences in the labour market, we aggregate the data to the one-digit SOC level, which consists of 22 major occupational categories. For each category, we compute the mean exposure and standard deviation across all occupations. Figure \ref{fig:AIexposureMeasures} displays these aggregated results, where the height of each bar represents the average exposure score, and the error bars denote the standard deviation. The colour of the bars reflects the logarithm of the mean wage, with lighter colours indicating higher wages.

The plot for the measure by \cite{Frey2017} indicates that the highest exposure levels are in "Office and Administrative Support" and "Installation, Maintenance, and Repair" occupations. The standard deviations are considerable, often reaching nearly half the height of the corresponding bar, suggesting substantial within-category variation. In contrast, the SML scores estimated by \cite{Brynjolfsson2018} are similarly high across all occupational categories, implying that exposure is widespread but mostly varies within categories rather than across them. The patent AI exposure measure developed by \cite{Webb2020} shows the highest levels of exposure for ``Business and Financial Operations," ``Computer and Mathematical Occupations," as well as ``Architecture and Engineering." Interestingly, this measure also highlights significant exposure for some lower-skilled occupations, such as ``Office and Administrative Support" and ``Production".

Measures of exposure, respectively to AI and GPT, in \cite{Felten2021} and \cite{Eloundou2023} show similar trends, with the highest exposure concentrated in higher-skilled occupations. Categories like ``Business and Financial Operations," ``Computer and Mathematical Occupations," and ``Education, Training, and Library" have elevated exposure levels. Among lower-skilled occupations, ``Sales and Related" and ``Office and Administrative Support" exhibit notable exposure as well.

The Mean Normalised Measure smooths out some of the differences observed in the individual metrics, identifying "Business and Financial Operations," "Computer and Mathematical Occupations," "Sales and Related," and "Office and Administrative Support" as the most exposed overall. This suggests that, across different methodologies, these occupational groups consistently emerge as being particularly vulnerable to AI-driven changes.

\subsubsection{Exposure rates in occupational networks.\label{comparison_net}}

To understand the impacts of AI across the labour market, it is crucial to consider labour mobility. This means that a static analysis of AI exposure of specific jobs looks only at one job at a time without taking into consideration which other job options are either available or perhaps also at risk for workers in an AI exposed job. Networks have emerged as a valuable tool for understanding these dynamics \citep{delRio2021,Moro2021}(Bocquet, 2024). In this analysis, we utilise the method developed by \cite{Mealy2018}, where occupations are linked based on the fraction of intermediate work activities they share, with each activity weighted according to its scarcity across all occupations. This approach creates clustered networks that reflect similarities across broad occupational categories, wages, and educational requirements, and also predicts potential job transitions.

\begin{figure}
    \centering
    \caption{US Occupation network\label{fig:OccupationalNetwork}}
    \includegraphics[width=0.45\linewidth]{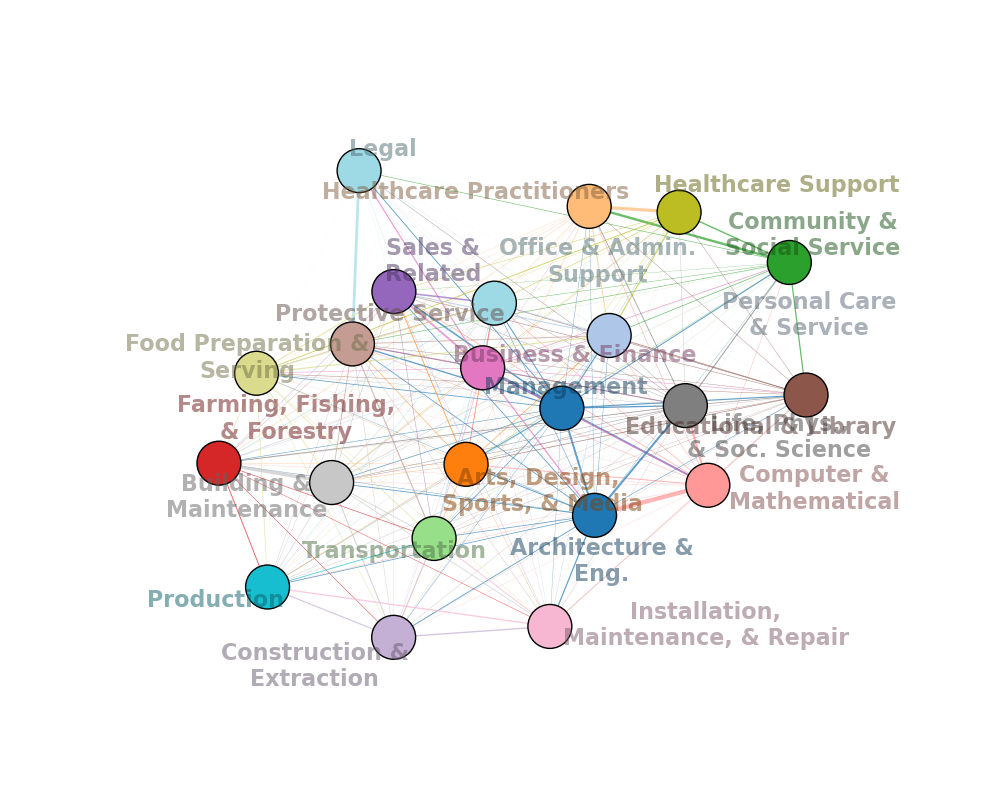}\includegraphics[width=0.45\linewidth]{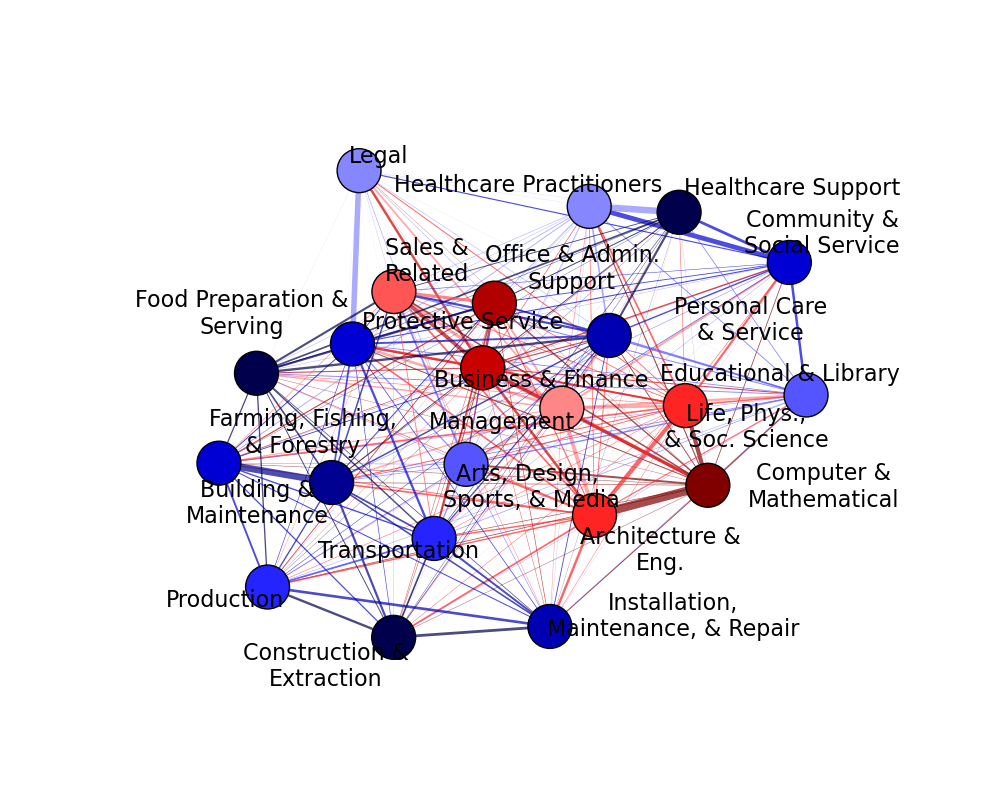}\includegraphics[height=0.25\linewidth]{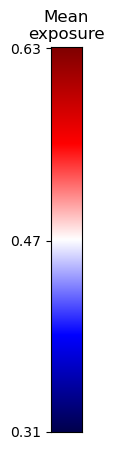}
    \pbox{15cm}{\footnotesize Network representation of occupation similarity based on the method developed by \cite{Mealy2018} of intermediate work activities. Nodes are occupations and edges denote the overlap of work activities. The left panel shows the network with labels and the right panel shows the network with nodes coloured by their exposure to automation.}
\end{figure}

Figure \ref{fig:OccupationalNetwork} presents the occupation network at the 1-digit level. On the left side of the figure, the network is coloured by broad occupational categories. We observe intuitive clustering patterns: occupations related to production, building and maintenance, and construction \& extraction are grouped together in the lower part of the network. Occupations centred around management and administrative support occupy the central region, while healthcare practitioners, support staff, community and social services, and personal care services are located towards the top of the network.

On the right side of figure \ref{fig:OccupationalNetwork}, the colour of each broad occupational category indicates their mean exposure to automation. The results show that the categories with the highest exposure—management, office and administrative support, and business and finance—are positioned centrally within the network. This central position might suggest that these occupations are highly interconnected within the broader labour market, raising several potential implications. On the one hand, affected workers might find it easier to transition into other roles due to the broader connectivity and versatility of their skill sets. On the other hand, significant job reductions in these key occupational clusters could disrupt existing pathways for occupational mobility, potentially leading to increased segregation in the labour market. Additionally, the interconnected nature of these occupations might mean that the impact of automation could have spillover effects, affecting not only directly exposed roles but also the jobs that depend on or interact closely with them.

\section{The impact of AI on jobs: Ex-post evidence\label{subsec:ex-post}}


While ex-ante analyses in the form of exposure measures provide useful insights into the potential impact of AI on employment, they are not sufficient to determine the overall impact of AI on labour markets, workers and firms. As discussed in Section
\ref{sec:theory}, several forces operate simultaneously and often in opposite directions—displacement of tasks, productivity gains and increasing demand, the creation of new tasks and reorganization of work processes. Whether AI is adopted by firms and organizations on a large scale, and what its effects on different workers and firms are, is ultimately an empirical question, shaped not only by the technology itself but also by its implementation in the broader economic context. 

Reflecting on this complexity, a fast-growing empirical literature focuses on the use of GenAI and reports a wide range of -- sometimes conflicting -- labour market effects. These divergent findings highlight the importance of context, including differences across occupations, sectors, institutional and national settings, in shaping the impact of AI on work. In this section, we review the main empirical findings on the impact of AI on productivity, on replacement of workers or tasks, and on work (re-)organization. To structure the discussion, we group studies according to the method that was used in the study as well as their scope. A growing number of papers examine the impact of GenAI on workers in experimental settings on the micro level (Subsection \ref{subsec:experiments}). Another set of studies, also mainly focusing on AI's effect on productivity, use natural experiments (Subsection \ref{subsec:nat_experiments}). Beyond micro-level productivity effects, we review broader labour demand and supply effects in Section \ref{sec_supply_demand}.



We capture the most recent literature using experiments by adopting a two-stage search protocol. First, we ran keyword searches in \href{https://scholar.google.com/}{Google Scholar}, \href{https://arxiv.org/}{arXiv}, \href{https://www.ssrn.com/index.cfm/en/}{SSRN}, and \href{https://www.nber.org/papers}{NBER Working Papers}, limiting results to English-language studies that used an experimental design to identify impacts of AI on employment, wages, productivity, or related outcomes. We use a combination of general keywords (``artificial intelligence'', ``generative ai'',``large language models'', ``ai and machine learning'', ``productivity'', ``performance'', ``labour markets'', ``wages'', ``employment'') and method and data specific search terms (``experiments'', ``RCT'',``field experiment'', ``natural experiment''). We limited our search to studies published from 2022 onward. After de-duplicating records and applying our inclusion criteria, we conducted backward snowballing—screening the reference lists of each focal study—to identify earlier but still relevant experimental papers. We then performed forward citation tracking to capture the most recent studies citing those seminal works, thereby ensuring comprehensive coverage of the latest contributions.

\subsection{Classification of task complexity}

An important insight from our review is that the effects of GenAI on work outcomes may depend on the complexity of the tasks to which it is applied. In line with theoretical arguments and prior work, we adopt a task-complexity perspective and classify the studies in our review accordingly, using the rubric outlined in Appendix \ref{appendix:task-complexity}. This classification was undertaken ex ante, based on conceptual considerations rather than observed outcomes, leaving open whether complexity ultimately matters for the effects of GenAI. Importantly, we do not treat entire occupations or activities as inherently “simple” or “complex.” For example, writing or programming can involve either simple or highly sophisticated tasks. Our classification reflects the context of each study and how the task under analysis maps onto our rubric.

Our classification builds on \cite{autor03}'s concept of routine tasks (those following explicit, codifiable rules), \cite{Acemoglu2025}'s distinction between easy-to-learn and hard-to-learn tasks, and collective-intelligence results on how diversity and interaction structure affect performance \citep{hong04,page08,almaatouq21,burton21}. To capture what determines effective AI assistance in practice, we identify four dimensions that shape whether workers can successfully leverage GenAI tools:

\begin{itemize}
\item \textbf{High-complexity tasks} score high on at least three of four dimensions: (1) knowledge requirements—demanding specialized expertise that cannot be easily extracted from training data (extending \cite{Acemoglu2025}'s hard-to-learn concept), (2) goal ambiguity—lacking clear success criteria or having subjective evaluation standards (updating \cite{autor03}'s routine/non-routine distinction, as AI can handle some non-routine tasks with clear goals), (3) task interdependence—involving critical dependencies following O-ring theory where component failure causes system-wide failure \cite{kremer93}. In such settings the returns to cognitive diversity and to the network structures that coordinate it are higher \citep{hong04,almaatouq21,burton21} and (4) resource/context demands—requiring extensive contextual information, specialized tools, or tacit knowledge accumulated through industry or professional experience that exceeds typical AI training data or context windows. Because this knowledge is dispersed, complementary expertise and deep-level diversity become especially valuable \citep{page08,cui24}. These tasks require iterative exploration and strategic prompting to extract value from AI tools.
    \item \textbf{Low-complexity tasks} meet fewer than three high-complexity criteria, meaning that in generally they are more codifiable\footnote{\cite{autor03} define "routine tasks" as rule-based and codifiable, which is overlapping with our understanding of "simple".} or well defined, with single-step outputs or straightforward procedures for how to reach a good solution (e.g., drafting generic emails, debugging simple code or answering well-structured questions). Low-complexity tasks tend to have a straightforward mapping from task goal to AI assistance: a typical user can immediately see 'what to ask' and obtains most of the value with little iteration or prompt engineering. 
\end{itemize}

To label each study according to transparent criteria we create a four-dimensional classification rubric including knowledge requirements (1), goal definition and success criteria (2), task interdependence (3), and resources and context requirements (4). For details on how we define each dimension, see Appendix \ref{appendix:task-complexity} and for examples of scores of different studies \ref{tab:example_class}.

\subsection{Productivity effects in experimental settings\label{subsec:experiments}}


In this section, we review the expanding body of randomized controlled trials (RCTs) that examine how GenAI (often ChatGPT) affects workers' productivity across differently skilled workers. In RCTs, workers are randomly assigned to either of two groups: a treatment group, encouraged to use GenAI, or to a control group, which has no access to GenAI.  Thanks to this random assignment, the difference in outcomes between the two groups can be interpreted as the causal effect of GenAI on worker performance. The RCTs include both laboratory experiments, conducted in highly controlled environments with recruited participants, and field experiments, implemented in real-world workplaces or platforms. Both are micro-level studies based on random assignment, but they differ in what they deliver: laboratory experiments offer strong internal validity by isolating mechanisms under controlled conditions, whereas field experiments provide greater external validity by capturing behaviour in actual work settings. Taken together, these approaches provide complementary insights into the micro-level effects of AI adoption.

Worker performance is measured based on either the time taken to complete a task and/or the quality of the output, assessed by independent evaluators. These experiments vary in two key ways. First, RCTs target different groups of workers—for example, consultants, web developers, or customer service agents. Second, they differ in the complexity of the tasks, ranging from simple editing or drafting exercises to more elaborate assignments designed to approximate real workplace activities. Yet, because RCTs are typically conducted in controlled, artificial settings, they often lack realism. As a result, their findings are highly context-specific and should not be generalized to broader labour-market dynamics or macroeconomic outcomes.




\subsubsection{RCTs in controlled settings\label{subsec:controlled_experiments}}

A first group of studies analyzes whether GenAI enhances productivity using RCTs conducted in laboratory settings—that is, in controlled, artificial environments. Worker performance is measured based on either the time taken to complete a task and/or the quality of the output, assessed by independent evaluators. These experiments vary in two key ways. First, they target different groups of workers—for example, consultants, web developers, or customer service agents. Second, they differ in the complexity of the tasks, ranging from simple editing or drafting exercises to more elaborate assignments designed to approximate real workplace activities. Yet, because these RCTs are conducted in controlled, artificial settings, they often lack realism. As a result, their findings are highly context-specific and should not be generalized to broader labour-market dynamics or macroeconomic outcomes.

Overall, these studies show that GenAI typically accelerates performance in tasks such as writing and coding, although improvements in quality are not guaranteed. The benefits depend on context and user skill, with less-skilled users gaining the most from well-defined tasks, while higher-skilled users tend to benefit more as tasks become complex and open-ended. In the following, we review RCTs beginning with those that examine relatively simple tasks according to our rubric and then turning to studies of more complex tasks, while noting that a sharp line cannot always be drawn for each individual study as a whole.


\paragraph{Simple Tasks} Most RCTs studies simple tasks. These studies focused on basic writing, some programming and mathematics tasks, debating and classification tasks.

\subparagraph{Writing.}
Across diverse writing domains -- professional, legal, and academic -- generative AI tools reliably increase writing speed, often substantially. In mid-level professional work, such as report writing or drafting delicate emails, \cite{Noy2023} report that participants using GenAI completed tasks about 40\% faster than those in the control group. For both argumentative essays and creative stories, \cite{li2024value} show that AI-assisted writers complete tasks more quickly and with greater ease. In a legal context, \cite{Choi2024Lawyering} find a 22\% average reduction in task completion time for law students using GPT-4 on realistic legal writing tasks, including drafting a complaint, a contract, a section of an employee handbook, and a client memo. Notably, these speed gains were widespread, benefiting both higher- and lower-performing students. 

While offering clear benefits for speed, the effect of GenAI use on quality is mixed. In creative writing tasks, \cite{doshi2024generative} describe a phenomenon they call “creative convergence,” where users tend to adopt similar AI-generated suggestions. This leads to less diverse, more homogenized content, even as the overall writing is rated as more enjoyable and creative—especially for less creative writers. Similarly, in the advertising domain, \cite{Chen2024CollaborationModality} find that AI-generated content can reduce originality. Their study shows that the “ghostwriter” mode—in which AI is used extensively—can lead to anchoring effects that reduce the output quality of higher-skilled creatives. In contrast, the “sounding board” mode, which provides feedback rather than generating content, tends to enhance quality for lower-skilled users without downsides for experts. \cite{li2024value} further show that the way users interact with GenAI—revising AI-generated content, or writing with AI suggestions—shapes outcomes in creativity and user satisfaction. While AI assistance can improve confidence and ease the writing process, users report that revising AI-generated drafts may lead to diminished creativity and a reduced sense of ownership. Together, these studies highlight that GenAI’s impact on work quality is context-dependent. It can elevate output—particularly for less experienced users—but may also constrain originality and shift users into more passive roles, depending on how exactly the AI tool is deployed. 

A consistent theme across multiple experimental findings is the disproportionate benefit GenAI provides to lower-skilled individuals, helping narrow performance gaps between workers of differing ability levels. In mid-level professional tasks, such as business analysis and strategy development, \cite{Noy2023} find that the most significant productivity gains were observed among lower-skilled workers, effectively bringing their performance up to par with their higher-skilled peers. Similarly, in legal writing, \cite{Choi2024Lawyering} show that while quality improvements from GPT-4 were modest overall, they were disproportionately beneficial to lower-performing law students. In creative writing, both \cite{Chen2024CollaborationModality} and \cite{doshi2024generative} find that AI elevates the performance of less creative individuals while having no or even a negative effect on the most creative participants. Yet a caveat is warranted. Much of the “leveling” in simple writing tasks may to come from what we term passive pasting—submitting AI output with little or no modification. In \cite{Noy2023} experiment, the majority treated participants submitted ChatGPT-generated text without modifying it or editing only slightly, with an average editing time of 3 minutes. This pattern may imply substitution rather than durable upskilling: when models already meet the bar on simple tasks, employers may use AI to automate.




\subparagraph{Programming and mathematics.} Evidence from programming tasks reinforces a now-familiar pattern: generative AI significantly boosts speed, delivers mixed or neutral effects on quality, and disproportionately benefits less-experienced users. In a pioneering study, \cite{Peng2023} examine the impact of GitHub Copilot -- an AI coding assistant -- on freelance web developers asked to implement an HTTP server in Java. The treatment group completed the task 55\% faster than the control group, despite no significant difference in task completion rates. Importantly, the productivity gains were especially large among older and less-experienced developers, suggesting that AI support is particularly valuable for those with weaker baseline skills. Unlike in some writing task experiments where the effect on quality was potentially negative for experts (e.g. \cite{Chen2024CollaborationModality}), the study by \cite{Peng2023} found no measurable decline in code quality, even as speed improved considerably. For narrow and well-scoped programming tasks GenAI seems to act as a skill equalizer. 

However, the pattern seems to shift when users face more complex problem-solving tasks. In a study of Korean university students, \cite{kim2024helping} evaluate the effects of ChatGPT on three competency areas: reading and writing, mathematical problem-solving, and computational thinking. While writing and computational thinking showed no significant changes overall—and modest benefits for intermediate students—ChatGPT assistance significantly decreased performance on the math task, particularly among low-ability participants who struggled to identify hallucinated or misleading AI responses. The key issue was misguided reliance: students with weaker math skills struggled to distinguish plausible but incorrect AI-generated answers from accurate ones. Lacking the ability to verify responses, they over-relied on faulty suggestions, leading to worse outcomes. These results challenge the notion that GenAI universally benefits lower-skilled users, revealing that assistance may be harmful when users lack the skills to critically assess AI-generated content.

\subparagraph{Classification tasks: radiology and portrait images.} Beyond writing and programming, researchers have tested the impact of AI tools on human performance in classification tasks, i.e. where users must label, categorize, or diagnose based on visual inputs. In a medical context, \cite{Agarwal2023} investigate how AI assistance affects radiologists diagnosing chest X-rays. Despite the AI outperforming roughly 75\% of individual radiologists on average, access to AI predictions did not yield overall improvements in diagnostic accuracy. Radiologists did respond to AI input -— adjusting their assessments in its direction -- but the integration proved suboptimal. Accuracy improved when the AI’s predictions were highly confident, but deteriorated when it was uncertain, exposing a vulnerability known as automation neglect: users underweight the AI’s input or fail to integrate it properly, especially when signals conflict. More broadly, the study reveals a fundamental coordination problem: radiologists and AI each possess valuable, partially unique information, but users struggle to combine these sources effectively. In fact, the majority of cases were best handled either by the radiologist or the AI alone, but not by their collaboration. Radiologists also took significantly longer to make decisions when AI input was available, suggesting cognitive overload or second-guessing.

A similar pattern emerges in simpler non-medical image classification. \cite{caplin23} study how individuals use AI support in a basic visual classification task: determining whether a person in a photo is over the age of 21. Despite the simplicity of the task compared to medical diagnostics, outcomes again depended heavily on user characteristics and belief updating. Less-skilled workers benefited more from AI assistance, echoing findings from programming and writing studies. However, improvements were not solely a matter of skill. Holding ability constant, individuals with more accurate self-assessments—those who better understood when they were likely to be right or wrong—gained the most from AI. This highlights the role of meta-cognitive awareness in effective human-AI collaboration. These findings suggest that performance in classification tasks hinges not just on task difficulty or baseline ability, but on users’ reflective ability to recognize when AI assistance is helpful—and when it is misleading—and to use it accordingly.

Table \ref{tab:rct-simple} provides an overview of RCT studies in laboratory settings that focus on relatively simple tasks.

\begin{footnotesize}
\rowcolors{2}{gray!15}{white}
\begin{longtable}{L{1.5cm}L{2.3cm}L{2.5cm} L{1.5cm} L{2.2cm} L{2.3cm}}
\caption{RCTs in controlled settings – Simple Tasks}\label{tab:rct-simple} \\
\toprule
\textbf{Study} & \textbf{Task} & \textbf{Sample \& AI Model} & \textbf{Completion Time} & \textbf{Quality} & \textbf{Inequality} \\
\midrule
\endfirsthead

\toprule
\textbf{Study} & \textbf{Task} & \textbf{Sample \& AI Model} & \textbf{Completion Time} & \textbf{Quality} & \textbf{Inequality} \\
\midrule
\endhead

Noy \& Zhang (2023) & Writing: Professional writing (press releases, short reports, analysis plans, and delicate emails; 20- to 30-min assignments) & 453 college- educated professionals; ChatGPT (GPT-3.5) & About 40\% faster with AI & About 18\% better grades with AI & Reduction: Initial inequalities halved in the treatment group \\
\addlinespace
Doshi \& Hauser (2023) & Writing: Creative writing (short eight sentence stories) & 293 participants recuited from Prolific; ChatGPT (GPT-4) & Not reported & Stories more creative but less unique and less diverse with AI & Reduction: Less creative writers benefit more \\
\addlinespace
Peng et al. (2023) & Programming: implement an HTTP server in JavaScript & 95 freelance programmers; GitHub Copilot (Codex) & About 55.8\% faster with AI & No measurable decline in code quality & Reduction: Developers with less programming experience, older programmers, and those who program more hours per day benefited the most \\
\addlinespace
Chen \& Chan (2024) & Writing: Advertisement copywriting (creating advertisements for an iPhone protective case) & 355 college- educated participants recruited from Prolific (marketing experts and non-experts); ChatGPT vs. control; GPT-4 & Not reported & Quality increases for non-experts; experts perform worse when using LLM as ghostwriter (anchoring effect and lower diversity of ads) & Reduction: quality gap narrows as non-experts improve \\
\addlinespace
Li et al. (2024) & Writing: Professional and creative writing (200-250 word article, either an argumentative essay or a creative story, within 45 minutes) & 379 participants from Prolific; ChatGPT (GPT-3.5) & Faster with AI support & Better language and fewer grammar mistakes but decreased satisfaction with writing experience and writing outcome, and a decline of diversity of writing & Not reported: Focus only on perception of usefulness not actual performance differences: High-confidence writers attach little value to AI-assistance in particular in creative writing \\
\addlinespace
Kim \& Moon (2024) & Mixed: reading and writing, mathematical problem-solving and computational thinking (20' each) & 349 Korean undergrads from different fields; ChatGPT (GPT-3.5) & Not reported & ChatGPT’s assistance significantly decreases the average score of the math task; quality of other tasks unchanged & Increase: low-ability participants harmed more \\
\addlinespace
Roldán-Monés (2024) & Debate preparation: competition of three to four rounds of short, one-on-one debates & 129 undergraduate students; ChatGPT (GPT-4) & Not reported - perception of having enough time increased in the treatment group & Treatment debaters are 9 \% more likely to win but no statistically significant & Increase: only top performers benefited from AI support \\
\addlinespace
Caplin et al. (2024) & Classification: Predicting age from portrait pictures (160 faces each) & 732 Prolific users; age estimation AI model 'Caffe' & Not reported & AI assistance improves overall prediction accuracy on average, but magnitude and direction of effect depends on whether participants are well calibrated about their own ability & Reduction: 34\% reduction in performance inequality from AI  assistance; most for well calibrated low-skill participants \\
\addlinespace
Agarwal et al. (2024) & Classification: Chest X-ray image diagnosis & 227 professional radiologists; CheXNet-style CNN supervised machine learning model & Slower with AI support & AI assistance does not improve humans’ diagnostic quality on average & Not addressed \\
\addlinespace
Choi et al. (2024) & Writing: Legal writing \& analysis: drafting a complaint, a contract, a section of an employee handbook, and a client memo & 60 law students; GPT-4 & About 22\% faster (average over the four tasks) & small in magnitude and inconsistent across tasks & Reduction: significantly more useful to lower-skilled participants \\

\bottomrule
\end{longtable}
\end{footnotesize}

\subparagraph{Debating.} As tasks grow more complex -- incorporating ambiguous objectives, open-ended exploration, and interdependent components -- the role of AI seems to shift to conditional augmentation of workers. Competitive debating offers a compelling test case. While it includes structured elements like argument writing, it also demands spontaneous verbal engagement, rhetorical agility, and social intuition, i.e., capabilities where generative AI is less directly useful. In a study of university debate teams, \cite{Roldan2024GenAIinequality} examine the effects of ChatGPT access on performance across students of varying skill levels. On average, the tool did not improve overall outcomes. However, it significantly benefited higher-skilled participants, who were better able to prompt the AI effectively and selectively incorporate its suggestions. In this setting, AI is not a substitute for core debating skills but a complement to them.

This pattern marks a departure from studies that focused on simpler tasks in the context of programming, writing or classification, where AI has equalized outcomes across skill levels. Instead, in debating, productivity gains from AI appear to amplify pre-existing differences. While lower-skilled students could use ChatGPT to improve clarity or structure in written preparation, they did not benefit in areas requiring judgment—such as distinguishing strong from weak arguments. Perceptions also diverged: high-skilled students reported a significantly greater sense of having had enough time to prepare when given access to ChatGPT, even relative to their lower-ability peers. These results illustrate that with growing task complexity, AI effectiveness depends on a user’s capacity to strategically engage with its output. As tasks begin to require more open-ended reasoning and integration of loosely structured inputs, AI becomes less of a universal productivity tool and more of a force multiplier for those already equipped with higher-order cognitive skills.

\paragraph{Complex Tasks\label{subsec:complexTasks}}

We identified four recent RCTs that evaluate the impact of generative AI on more complex tasks, as defined by our labeling rubric. These studies span academic writing \citep{usdan2024generative}, legal reasoning \citep{Schwarcz25}, general knowledge work \citep{Haslberger2025NoEqualizer}, and high-end professional consulting tasks \citep{dellacqua23} involving varied degrees of complexity and ambiguity. We summarize these studies in Table \ref{tab:rct-complex}. Across all four studies, generative AI tools led to clear improvements in speed and generally positive effects on output quality. However, inequality dynamics are mixed: while some evidence suggests a leveling effect, other findings imply persisting or even increasing inequality, particularly in more ambiguous or judgment-heavy tasks.

\subparagraph{Academic writing.} In an academic writing context, \cite{usdan2024generative} examine how graduate students use ChatGPT to complete professional memo assignments. Students using the tool reduced their writing time by 64.5\%, and their average scores rose from a B+ to an A. The largest relative gains were observed among students for whom English is a second language (ESL), helping to close pre-existing performance gaps in both speed and quality. However, the study only measures inequality in terms of language background, not writing ability more broadly. This limits interpretation, as the observed ``leveling effect'' reflects language support for non-native speakers rather than a clear difference between high- and low-skill writers—making the equity gains somewhat expected. Beyond performance outcomes, the study attempts to shed light on how students interacted with the AI—whether they used it as a substitute for effort or a collaborative partner in the writing process. While the study lacks detailed behavioral tracking, qualitative feedback indicates that students did not simply paste in AI outputs, but refined, reorganized, and combined AI-generated content with their own input leaving judgment and argumentation under human control.

\subparagraph{Legal writing.}A second study, focusing on legal writing and reasoning, conducted by \cite{Schwarcz25}, compares the effects of two distinct AI tools on student performance across six realistic legal tasks. The authors evaluate Vincent AI, a retrieval-augmented generation (RAG) model tailored to legal content, and o1-preview, a general-purpose reasoning model developed by OpenAI. Both tools significantly increased speed, with time savings ranging from 12\% to 37\% depending on the task and model. However, the tools differed in how they supported users. Vincent AI primarily boosted speed, especially for lower-skilled students, but had more modest effects on quality.

In contrast, o1-preview offered stronger quality improvements—particularly in analytical depth and clarity—while also delivering modest speed gains. These quality gains were disproportionately concentrated among lower-skilled users, suggesting that o1-preview effectively raised the performance floor. Interestingly, the study finds that performance increase with AI was strongest on the most complex but clearly structured legal task. In the more ambiguous assignment, outcomes varied more widely. For some high-performing students the authors report performance declines, a phenomenon they describe as a “falling asleep at the wheel” effect, where passive use of AI led to reduced cognitive engagement. Behavioral and survey data point to a mix of augmentation and substitution dynamics: some students used AI to extend their reasoning and improve structure, while others appeared to offload effort without sufficient critical engagement. Overall, the study shows that the effects of AI depend not just on who is using it, but on what kind of tool is used, how it fits the task at hand, and whether users actively engage with its suggestions or default to passive reliance on AI output.

\subparagraph{Consulting.} The third study by \cite{dellacqua23} examines the effects of AI on high-end professional consulting tasks, with a particular focus on how users navigate the uneven frontier of AI’s capabilities. The authors conduct a RCT involving over 758 management consultants performing complex tasks designed to mirror real-world workflows. Their central contribution lies in identifying a ``jagged technological frontier'': a boundary where GPT-4 excels at certain tasks but falters at others of comparable human difficulty. In the experiment, consultants were randomly assigned to perform either a qualitative creative task—developing and marketing a new product—or a quantitative strategic task requiring interpretation of mixed data sources. For first task, which fell within GPT-4’s capability frontier, AI assistance significantly improved performance: consultants completed more tasks, produced higher-quality outputs, and worked faster. In this context, high ``retainment'' of GPT-4 output—i.e., copying AI-generated text with minimal edits—was positively associated with better performance, indicating that strategic delegation to the AI was effective. However, the use of GPT-4 also led to less diverse outputs across participants, echoing findings from simpler tasks that generative AI tends to increase average quality while reducing variation in responses. Lower-performing consultants benefited most, suggesting AI’s potential to equalize skill differences in high-end knowledge work.

In contrast, for the second task—outside the AI frontier—AI assistance reduced accuracy of responses (a binary recommendation) by 19 percentage points, even as it improved perceived quality and reduced completion time. Here, blind reliance on AI outputs (“high retainment”) was associated with poorer performance, echoing previous findings that blind trust in AI systems, in particular in ambiguous open-ended settings, can degrade performance because humans seem to struggle to critically recognize the limits of AI output. Conversely, consultants who interrogated, amended, or selectively integrated AI suggestions performed better, underscoring the importance of skillful human-AI collaboration in complex decision-making. Two successful strategies emerged fro this context: The first, ``Centaur'' behavior, involved a strategic division of labor: consultants delegated specific sub-tasks to GPT-4 (such as summarizing or drafting) while retaining full control over areas requiring judgment, contextual interpretation, or domain expertise. This model relied on users’ ability to assess the relative strengths of human and machine across task components and to consciously allocate responsibility accordingly. The second, ``Cyborg'' behavior, reflected a deeper form of integration, where users interacted with the AI at a granular level—editing its outputs, prompting iteratively, or even co-authoring responses line by line. Rather than dividing labor, Cyborg users treated AI as a cognitive partner, often re-prompting or refining suggestions in real time. Crucially, both strategies—despite their differences—enabled consultants to navigate the jagged frontier more effectively than those who engaged passively with the tool. Yet the study also emphasizes that such frontier navigation is inherently difficult: even experienced professionals, working on tasks closely related to their daily roles, often failed to recognize when AI was likely to mislead. This suggests that realizing the full value of generative AI in professional contexts requires not only technical access but also meta-cognitive skill, judgment, and an evolving understanding of AI’s limitations. Integration style is thus not a neutral design choice, but a determinant of whether AI augments or undermines expert performance.

\subparagraph{General knowledge work.} The last study in this group takes a broader view, examining how generative AI affects performance across a range of general knowledge tasks, administered to a near-representative sample of the UK working-age population spanning multiple occupations. In this large-scale online experiment, \cite{Haslberger2025NoEqualizer} study the effects of ChatGPT across three tasks—email editing, a policy assessment, and a comprehension task. The tasks were deliberately designed to vary in complexity, enabling a within-study comparisons of how AI mediates performance across simple and complex tasks. The authors find that generative AI improves both speed and quality across all tasks, with the largest benefits observed for the complex but unambiguous comprehension task. However, the study finds little evidence that ChatGPT narrows performance gaps across demographic groups; in fact, inequality increased slightly in the more ambiguous assessment task, particularly among those with less experience using AI. Thus, ChatGPT did not reduce performance differentials between educational or occupational groups, and inequalities between younger and older workers even increased. These results underscore once again that AI effectiveness depends not only on task complexity but also on task clarity, with ambiguous tasks requiring greater prompting skill and familiarity with AI.

\begin{footnotesize}
\rowcolors{2}{gray!15}{white}
\begin{longtable}{L{1.5cm} L{2.3cm} L{2.5cm} L{2.0cm} L{2.5cm} L{2.5cm}}
\caption{RCTs in controlled settings – Complex Tasks}\label{tab:rct-complex} \\
\toprule
\textbf{Study} & \textbf{Task} & \textbf{Sample \& AI Model} & \textbf{Completion Time} & \textbf{Quality} & \textbf{Inequality} \\
\midrule
\endfirsthead

\toprule
\textbf{Study} & \textbf{Task} & \textbf{Sample \& AI Model} & \textbf{Completion Time} & \textbf{Quality} & \textbf{Inequality} \\
\midrule
\endhead

Dell’Acqua et al. (2023) & Consulting (set of realistic management consulting tasks: inside and outside of frontier of AI capabilities) & 758 consultants; GPT-4 & Inside frontier: 23\% -- 28\% faster; Outside frontier: 18\% -- 30\% faster & Inside frontier: on average 40\% higher quality scores with AI compared to control; Outside frontier: increase in quality of recommendation but decrease of 19 percentage points in correctness & Reduction inside frontier: below-average performers gained 43\% vs. 17\% for above-average; No reduction on tasks outside the AI capability frontier \\
\addlinespace
Usdan et al. (2024) & Academic writing (summarize and assess scientific article, propose policy) & 27 graduate students; ChatGPT & Writing time reduced by 64.5\% & Quality rose from B+ to A (about 18\% increase) & Reduction: largest improvements for English as a Second Language (ESL) students, narrowing performance gap (speed and quality) \\
\addlinespace
Haslberger et al. (2025) & Various realistic work tasks; from low to high complexity: email, policy assessment and text comprehension task & 1041 survey respondents, UK working-age population; ChatGPT & Speed increased across tasks: time savings in seconds: 134 for emails, 235 for assessment, 101 for comprehension & Improvements across all tasks. Largest improvements in comprehension (complex but straightforward), lowest in assessment task (highest ambiguity) & Mixed: small increase in inequality in most ambiguous task \\
\addlinespace
Schwarcz et al. (2025) & Six realistic legal drafting \& research tasks (email, research memo, complaint analysis, nondisclosure agreement, motion, letter) & 127 upper‑level law students; Vincent AI (legal RAG) or OpenAI o1‑preview reasoning model & Vincent AI 14--37\% faster; o1‑preview 12--28\% faster vs. control & Both tools increased quality on most tasks (up to +28\% scores); Vincent AI least hallucinations but o1 larger gains in quality & Reduction: o1-preview helped reduce quality gap; Vincent AI helped reduce speed gap \\
\bottomrule
\end{longtable}
\end{footnotesize}

\subparagraph{Discussion: RCTs in controlled settings.} 

Most of experimental studies in controlled settings focus on whether GenAI narrows or widens the performance gap between high- and low-skilled workers, and report mixed results. A first wave of RCTs finds large positive effects of AI on productivity, mostly driven by low-skilled workers catching up with high-skilled workers — suggesting a reduction in inequality. A more recent wave of studies, finds smaller, often statistically insignificant average effects, with gains concentrated among high-skilled workers and even losses for low-skilled ones—indicating a potential widening of inequality. To rationalize this, researchers highlight the role of task complexity: AI support tends to reduce inequality in simple tasks but increase it in complex ones. We structure our analysis accordingly classifying experimental studies into low and high complexity based on task requirements. 

The majority of RCTs that we identified focuses on relatively simple tasks. Of the 18 studies examined, only five concentrate on complex tasks in their experimental designs. The simple tasks in the studies cover a broad spectrum of domains, ranging from general activities such as writing, customer support, and programming to more specialized tasks like x-ray diagnostics and legal practice. Overall, while there is a substantial body of research on simple tasks, studies focusing on complex, real-world tasks remain scarce.

Overall, evidence from RCTs using simple tasks suggests that GenAI consistently enhances performance in relatively simple, well-bounded tasks—particularly through time savings. In writing and coding, GenAI tools reliably speed up task completion, often by 20–55\% (\cite{Choi2024Lawyering, li2024value, Noy2023, Peng2023}, with few tradeoffs in accuracy or coherence. However, these efficiency gains are less pronounced—or entirely absent—in classification tasks, where human-AI collaboration seems less straight-forward (\cite{Agarwal2023,caplin23}).

Quality outcomes are mixed. In some contexts—such as routine writing—AI assistance improves output (\cite{Noy2023}). In others, particularly those involving creativity, it can flatten variation, introduce errors, or degrade originality (\cite{Chen2024CollaborationModality, doshi2024generative, kim2024helping}). Similarly, the distributional effects of AI vary across tasks. Several studies point to a narrowing of performance gaps, with lower-skilled individuals benefiting disproportionately from AI support (\cite{caplin23, Chen2024CollaborationModality, Choi2024Lawyering, doshi2024generative, Noy2023, Peng2023}). But this pattern is context-dependent: in tasks that require users to critically assess or strategically adapt AI output—such as math problem-solving (\cite{kim2024helping}) or debate preparation—higher (\cite{Roldan2024GenAIinequality})-skilled individuals are more likely to benefit, and inequality may widen.

RCTs using simple tasks come with important limitations. Most studies are conducted in controlled, artificial settings with narrow task scopes and short time horizons. As \cite{Noy2023} themselves note, these designs cannot capture the full dynamics of AI adoption in real-world settings—where tasks are more complex, collaboration is embedded in institutional structures, and skill development unfolds over time. In low-stakes lab tasks, GenAI may appear to increase productivity and equalize performance. But in real-world environments that require judgment, coordination, and constant adaptation and learning, the role of AI—and who benefits from it—may look different.

RCTs involving more complex tasks show that generative AI can improve both the speed and quality of work on complex tasks, echoing patterns seen in simpler ones. Productivity gains are common, especially for lower performers, but effects on inequality are mixed. In some cases, performance gaps shrink; in others—especially when tasks are ambiguous or judgment-based—they persist or widen. Across studies, task ambiguity stands out more than complexity as a key moderator of AI’s effectiveness. Structured tasks see the most consistent gains, while ambiguous ones produce more variable outcomes, depending on users’ AI skills and strategies. Although participants were generally encouraged to collaborate with the AI, evidence shows both active integration and passive over-reliance. The concept of a jagged technological frontier introduced by \cite{dellacqua23} offers a useful lens here: it suggests that the divergent outcomes observed in \cite{Schwarcz25} may reflect underlying differences in the capability frontiers of the two AI models tested—one better aligned with the task demands than the other. As with RCTs on simple tasks, the controlled lab-like design of these experiments offers valuable causal insights but also comes with limitations. The artificial nature of the setting, the relatively short time-frames, and the absence of institutional or team-based dynamics limit the possibility to generalise these findings to actual work environments. In the next section, we turn to field studies that examine how generative AI is being used in real-world workplace contexts, where organizational structures, norms, and incentives shape both adoption and impact.

These findings raise a broader question about whether AI acts as a substitute for human effort or a complement to human skill. In the simplest settings, the former often seems more accurate. For example, \cite{Noy2023} find that over two-thirds of participants submitted ChatGPT-generated content without modification. Rather than engaging in a collaborative process of revision or ideation, many users simply pasted AI responses, suggesting that productivity gains may reflect task outsourcing more than skill enhancement. This interpretation aligns with other findings, such as user-reported declines in creativity and sense of ownership when revising AI drafts. In these cases, GenAI seems to replace rather than augment human effort.

\subsubsection{RCTs in field experiments \label{subsec:field_experiments}}

A key limitation of RCTs in controlled laboratory environments is that participants operate under artificial conditions, raising questions about how far the results can be generalized to real-world work. To address this gap, such studies need to be complemented by experimental designs that capture the complexity and realism of actual workplace settings. Although more difficult to design, implement, and finance, field experiments hold greater promise for generating empirically meaningful insights. Natural experiments can partly fill this role, but they face two constraints: real-world work often blends simple and complex tasks, making categorization ambiguous, and researchers have far less control over the setting, context, and the precise nature of the tasks involved.


\paragraph{Simple Tasks}

While RCTs in lab settings offer strong internal validity, field experiments provide complementary insights by capturing real-world conditions of AI adoption—over longer time-frames, within complex workflows, and under looser constraints on tool use. We identified three such studies that focused on relatively simple tasks: two field experiments conducted within single firms -- one in telemarketing (\cite{jia2024and}) and one in customer support (\cite{Brynjolfsson2025}) -- and a broader study of general knowledge workers across dozens of companies from different industries (\cite{dillon2025shifting}). The evidence is mixed. While all three studies document productivity gains, they diverge in who benefits and how, underscoring that the context and specifics of AI implementation shape both outcomes and equity effects.

\textbf{Telemarketing}. \cite{jia2024and} study how AI can augment employee performance through a sequential division of labor in a telemarketing setting. In their field experiment, 3,144 customers were randomly assigned to 40 human sales agents, with or without assistance from a custom-built AI chatbot. The AI handled the initial, repetitive task of generating sales leads, while humans managed the subsequent sales persuasion stage. While AI assistance increased task efficiency by offloading routine work, low-skilled workers experienced slower completion times because the AI filtered out uninterested customers, leaving them with more difficult cases. In contrast, high-skilled agents maintained or improved speed while focusing on more substantive interactions. It seems high-skilled workers could better leverage the freed-up cognitive resources and complex interactions. The study documents a clear inequality effect, with AI assistance benefiting skilled employees disproportionately. Lower-skilled agents reported stress and reduced morale, while higher-skilled agents reported increased motivation, freedom, and satisfaction.

\textbf{Customer support}. In contrast, \cite{Brynjolfsson2025} study a very different human-AI configuration: real-time, conversational collaboration between customer support agents and a conversational chatbot. The authors find that AI access increased productivity by 15\%, with the largest gains—up to 30\%—seen among less-experienced agents. The AI helped novice workers mirror top performers, improving fluency, response speed, and even customer sentiment. Meanwhile, high-skilled agents saw negligible productivity improvement and, in some cases, a slight decline in interaction quality. This suggests that when AI acts more like a co-pilot than a background filter—offering direct, moment-to-moment assistance—it serves as a skill equalizer. Importantly, the quality of uptake mattered: agents who more frequently followed the AI’s suggestions learned faster, even during periods when the tool was unavailable. On average, adherence rates—measured as the proportion of messages based on AI suggestions—were around 38\% and similar across skill levels. However, the study does not break down whether agents modified or directly copied suggestions, nor does it analyze adherence behavior by skill group. This limits interpretation: it remains unclear whether the small declines in performance among high-skill agents stemmed from over-reliance on AI or from failure to adapt its output to their own expertise.

What appears as a contradiction between these two studies may, in fact, reflect a deeper principle: the structure of AI integration shapes who benefits. When AI handles the front-end work and delegates more complex tasks to humans (as in the sales example), it tends to amplify the advantages of skilled workers. But when AI works alongside users in a tightly coupled, interactive way (as in the customer service example), it can support lower-skilled users by reducing cognitive load and guiding decision-making in real time. Put differently, division of labor benefits experts; real-time side-by-side assistance benefits novices. More specifically, the sales setting illustrates augmentation: skilled workers gain because they are freed to focus on higher-value tasks that demand judgment and creativity. The customer support case reflects automation of human effort: here, the AI effectively performs part of the task itself, allowing novices to improve speed and quality by relying on ready-made suggestions. These contrasting outcomes underscore that the way AI is embedded in workflows—not just the task itself—shapes whether it substitutes for effort or amplifies skill. But they also raise a longer-term concern: if AI tools increasingly automate the core functions of low-complexity jobs, the role of humans may shift from being augmented by AI to merely supervising or refining AI outputs—or risk being replaced entirely. In such settings, it may be more accurate to describe the system as AI augmented by humans rather than the reverse.

\textbf{General knowledge work}. While the previous two studies offer detailed insight into effects within single firms, \cite{dillon2025shifting} take a broader lens, examining AI use across a large and diverse set of knowledge workers. Covering over 7,000 employees in 66 companies, the study tracks the early adoption of Microsoft 365 Copilot, a GPT-4-based tool. This broader scope comes at the cost of analytical depth: the study does not evaluate output quality, task complexity, or differential effects across skill levels or demographics. Instead, its core contribution is behavioral—showing that workers adopt AI tools when available and use them to save time on simple, individual-level tasks. Regular users spent 3.6 fewer hours per week on email—a 31\% reduction—and completed collaborative documents somewhat faster. However, total document output remained unchanged, and effects on meetings or team-based work were minimal. Even in teams where more coworkers had access to Copilot, effects remained minimal, suggesting that significant shifts in work practices require broader firm-level strategies. These findings underscore that while AI can help streamline individual-level routine work, more transformative productivity or equity effects may depend on widespread organizational adoption and more integrated team-level use. Moreover, the study illustrates how difficult it is to rigorously evaluate the impact of generative AI tools on worker outcomes at scale, beyond simple, speed-related productivity measures.

\begin{footnotesize}
\rowcolors{2}{gray!15}{white}
\begin{longtable}{L{1.5cm} L{2.3cm} L{2.5cm} L{2.0cm} L{2.5cm} L{2.5cm}}
\caption{Field Experiments – Simple Tasks} \\
\toprule
\textbf{Study} & \textbf{Task} & \textbf{Sample \& AI Model} & \textbf{Completion Time} & \textbf{Quality} & \textbf{Inequality} \\
\midrule
\endfirsthead

\toprule
\textbf{Study} & \textbf{Task} & \textbf{Sample \& AI Model} & \textbf{Completion Time} & \textbf{Quality} & \textbf{Inequality} \\
\midrule
\endhead

Jia et al. (2024) & Telemarketing: sequential division of labor, AI generates sales leads, humans do sales persuasion & 3,144 customers served by 40 sales agents, custom-made AI chatbot & Lower speed for low-skill workers because AI-preselected customers are more challenging to serve & Increased creativity and increased sales for high-skilled workers & Increase: positive effect for higher-skilled employees \\
\addlinespace
Brynjolfsson et al. (2025) & Customer support (service inquiries) & 5,179 customer support agents; GPT-3 & 15\% more issues resolved per hour, 30\% for novice workers & Improved customer satisfaction (sentiment ratings) for lower-skilled workers; small quality decline for the most experienced workers & Reduction: novice workers benefited the most, narrowing performance gap \\
\addlinespace
Dillon et al. (2025) & General knowledge work (emails, documents, meetings) & 7,137 workers across 66 large firms; M365 Copilot (GPT-4) & 31\% less time on email, 5--25\% faster on documents & Not reported & Not reported \\
\bottomrule
\end{longtable}
\end{footnotesize}


\paragraph{Complex Tasks}

Evidence from RCTs discussed previously suggests that complex, less structured tasks can pose challenges for workers to use AI effectively. We review two field experiments involving more complex tasks according to our classification: one with entrepreneurs in Kenya \cite{otis23}, and another with software developers at large firms \cite{Cui2025HighSkilled}. Both studies highlight substantial heterogeneity in outcomes, with individual skill levels and usage patterns playing an important role in determining AI’s effects.

\textbf{Entrepreneurship}. \cite{otis23} study the productivity effects of AI on entrepreneurs operating in unstructured environments, using a randomized experiment among small business owners in Kenya. Participants received access to an AI assistant via WhatsApp. On average, they do not find significant treatment effects. However, the effects are strongly heterogeneous: high-performing entrepreneurs increased their revenues by about 15 per cent, while low-performing entrepreneurs experienced an 8 per cent decline. To better understand these divergent outcomes, the authors conduct an exploratory text analysis of the WhatsApp chats between entrepreneurs and the AI assistant, along with survey responses describing business changes. Interestingly, the performance effects are not explained by differences in the frequency or types of questions asked, the nature of the AI’s suggestions, or the overall likelihood of implementing advice. Instead, the critical distinction lies in how entrepreneurs selected and applied the AI-generated suggestions. Low-performing entrepreneurs tended to implement generic advice—such as lowering prices or increasing advertising—which, when misaligned with their specific business context, often led to reduced revenues and higher costs. High performers, by contrast, generated more precise, context-specific questions to identify tailored and actionable changes, such as switching to a more effective car wash detergent or finding alternative power sources during blackouts. These findings highlight a strong complementarity between worker skill and AI effectiveness in unstructured environments. While the AI system made valuable advice more accessible to all users, only the more skilled entrepreneurs were able to effectively interpret and act on that advice. This echoes patterns observed in earlier studies discussed in our review—such as those in consulting (\cite{dellacqua23}), debating (\cite{Roldan2024GenAIinequality}), and telemarketing (\cite{jia2024and})—where AI support under amplified performance gaps, benefiting high performers while offering less or even negative returns to lower performers.

\textbf{Software development}. In a rare large-scale field experiment within high-skill, high-wage occupations, \cite{Cui2025HighSkilled} draw on experiments happening in three major firms—Microsoft, Accenture, and a large manufacturing company—to evaluate the effects of generative AI on software developers in real-world workplace settings. They find that access to AI tools leads to a 26 per cent increase in the number of completed tasks. Less experienced developers adopted the tools more readily and saw greater productivity gains, suggesting that generative AI can help narrow performance gaps within high-skilled occupations. The developers in these firms worked on a range of tasks, from routine to complex, and the allocation of work reflected their skill level and tenure. This setting thus provides a realistic view of how AI is used within high-paying, and highly exposed professions. The observed productivity gains among lower-skill developers mirror findings from earlier studies in customer support \cite{Brynjolfsson2025}, where AI tools offered the greatest relative uplift to less experienced workers. At the same time, the study’s broad scope—spanning three distinct firms—presents challenges. While it offers valuable external validity and broad coverage, this heterogeneity also limits the depth of insight into how AI adoption plays out in specific organizational contexts. As with \cite{dillon2025shifting}, differences in task composition, tool integration, and workplace practices across firms highlight the challenges of rigorously evaluating AI’s impact on worker outcomes at scale—particularly beyond surface-level metrics such as speed or task volume.

\begin{footnotesize}
\rowcolors{2}{gray!15}{white}
\begin{longtable}{L{1.3cm} L{2.1cm} L{2.4cm} L{2.5cm} L{2.4cm} L{2.4cm}}
\caption{Field Experiments – Complex Tasks} \\
\toprule
\textbf{Study} & \textbf{Task} & \textbf{Sample \& AI Model} & \textbf{Completion Time} & \textbf{Quality} & \textbf{Inequality} \\
\midrule
\endfirsthead

\toprule
\textbf{Study} & \textbf{Task} & \textbf{Sample \& AI Model} & \textbf{Completion Time} & \textbf{Quality} & \textbf{Inequality} \\
\midrule
\endhead

Otis et al. (2023) & Entrepreneurship in Kenya & 640 Kenyan entrepreneurs; GPT-4 via WhatsApp & Not reported & Overall no effect; high performers: +15\% revenue; low performers: --8\% revenue & Increase: high performers benefited, low performers negatively impacted by AI support \\
\addlinespace
Cui et al. (2025) & Software development (diverse enterprise-level coding tasks) & 4,867 developers (Microsoft, Accenture, Fortune 100); GitHub Copilot & Increase of 26\% in weekly number of completed tasks by developer; 14\% increase in code updates and 38\% increase in times code compiled & No quality loss reported (output maintained) & Reduction: less-experienced gained more, narrowing experience gaps \\
\bottomrule
\end{longtable}
\end{footnotesize}

\paragraph{Summary - Experimental Studies}



To summarise, the experimental evidence demonstrates very heterogeneous productivity gains from using AI tools. For simple tasks, the literature reports substantial productivity enhancements from GenAI assistance. These improvements are particularly pronounced among lower-skilled or less-experienced workers, suggesting that AI can reduce inequality in the workplace by improving the performance of lower-skilled workers.

As task complexity increases, however, results vary significantly. In complex tasks requiring nuanced judgment and specialized expertise, AI integration may lead to more frequent errors \citep{dellacqua23}. Studies have shown that professionals may over-rely on AI outputs or fail to optimally integrate AI assistance with their expertise, a phenomenon referred to as "automation neglect" \citep{Agarwal2023}. This indicates that AI's efficacy is highly task-dependent and that indiscriminate application of AI tools may not yield the desired improvements in productivity or accuracy.

\subsubsection{Productivity effects from natural (quasi-) experiments \label{subsec:nat_experiments}}

In contrast to experiments with RCTs, whether in controlled setting or as field experiments, natural experiments occur in authentic, real-world contexts, offering greater realism but sacrificing researcher control. Natural experiments exploit external events or situations that naturally resemble random assignment, such as abrupt policy changes, technological innovations, or unforeseen environmental factors. Since assignment to treatment or control conditions is effectively random, observed differences in outcomes between groups can be attributed to the intervention or event (e.g., access to AI tools). While natural experiments provide more realistic insights, researchers often lack detailed data on worker performance at the task level, which makes it more challenging to fully understand the causal mechanisms at play. 

\paragraph{Github - Programming} Programming has become a frequent setting for such studies, as online public repositories of collaborative coding platforms like GitHub provide rich, high-frequency and publicly accessible data on developer activity that can be leveraged for natural experiments. Building on this data, recent research has begun to examine how AI-based coding assistants—such as GitHub Copilot and ChatGPT—affect programming productivity and the organization of coding work. A growing number of studies exploit large-scale repository data and quasi-experimental variation to identify these effects. Findings point to modest productivity improvements overall, but also reveal substantial heterogeneity across developer experience levels, along with shifts in task allocation and innovation dynamics.

One early study by \cite{Kreitmeir2024Heterogeneous} exploits the temporary ban of ChatGPT in Italy in April 2023 to estimate the short-term effects of removing AI access. Using data from 36,000 GitHub users, they find no overall changes in productivity among experienced developers, but minor slowdowns in simple tasks like debugging. In contrast, less experienced developers increased both the quantity and quality of their output in the absence of AI, suggesting that ChatGPT may have reduced their performance. These findings align with experimental evidence suggesting that generative AI does not necessarily boost the productivity of junior developers and may, in some cases, even hinder it. However, the study has some limitations: the ban was relatively short-lived, not all developers may have been active users of ChatGPT prior to its removal, and some may have circumvented access restrictions through VPNs (Virtual Private Network). 

Several other studies focus on the introduction of GitHub Copilot, an AI coding assistant integrated directly into the GitHub platform. \cite{Song2023Copilot} study 4.9 million repository-month observation pairs from open-source projects on Github and find a 5.9 percent increase in code contributions following Copilot adoption, accompanied by an 8 percent increase in coordination time. While the increase in output reflects higher developer activity, it also comes with greater coordination costs, as evidenced by more extensive discussions and a larger number of contributors involved in each decision. In particular core developers—who are more familiar with the project—benefit from the tool, while peripheral developers benefit less and generate higher coordination costs. These results suggest that, beyond programming experience, project-specific context familiarity plays a relevant role in determining the effectiveness of AI tool use. \cite{Yeverechyahu24} further document that the increase in contributions driven by Github Copilot is concentrated in maintenance and refinement tasks, rather than novel feature creation. Their findings imply a shift in the innovation trajectory of software projects, with AI disproportionately supporting iterative improvements over more innovative capability-expanding work.

Beyond productivity, \cite{Hoffmann2025NatureOfWork} emphasize changes in the time allocation across workplace activities. They show that Copilot access leads developers to reallocate time from project management activities toward core coding. This shift is particularly pronounced among lower-ability programmers, who show greater increases in autonomous behavior and exploratory coding. AI thus appears to reinforce technical engagement while reducing the time spent on collaborative or organizational tasks. This stands in contrast to the findings of \cite{Song2023Copilot}, who document increased coordination efforts following increased coding activity after AI adoption. 

Finally, \cite{daniotti2025using} focus not on evaluating productivity effects directly, but on developing an innovative method to measure generative AI adoption at scale. Using a machine learning model trained to detect code written by large language models, they analyze over 80 million commits to open-source Python repositories between 2019 and 2024. Their approach allows them to estimate the share of AI-generated code contributions over time and across users. By the end of 2024, they estimate that approximately 30 percent of functions contributed by U.S.-based developers were written with the aid of AI. Leveraging variation in AI use at the developer level, they estimate that generative AI is associated with a 2.4 percent increase in output, measured in the number of commits. They also find that adoption is significantly higher among less experienced developers and in certain geographic regions, with substantial between-country variation. These patterns suggest that generative AI may be helping junior developers expand into new areas of programming and that, over time, such tools could contribute to narrowing experience-based productivity gaps, although cross-country variation in uptake remains large.

The studies examining GitHub coding platform offer complementary insights to the prior experimental evidence. While RCT studies have demonstrated substantial productivity gains in routine programming tasks—particularly for less-experienced workers—the natural experiments discussed here paint a more complex picture of AI’s effects in real-world programming environments. First, they confirm that generative AI tools such as GitHub Copilot are being rapidly adopted, especially by junior developers, and that they are associated with measurable increases in output. However, these gains are typically smaller than those observed in experimental settings and can be accompanied by increased coordination costs, shifts in the organization of work, or a reduction in innovative contributions. Several studies underscore that AI adoption may amplify productivity gaps, echoing experimental findings on the importance of contextual knowledge. Notably, the study by \cite{Kreitmeir2024Heterogeneous} finds that removing AI access improved the performance of junior developers, highlighting the risk that premature or uncritical adoption may lead to over-reliance and suboptimal performance. Together, these studies indicate that AI tools have started to reshape how programming work is structured—shifting contributions toward refinement, changing how teams coordinate, and altering the allocation of effort across tasks.

\begin{footnotesize}
\rowcolors{2}{gray!15}{white}
\begin{longtable}{L{1.3cm} L{2.1cm} L{2.4cm} L{2.5cm} L{2.4cm} L{2.4cm}}
\caption{Software Development - Natural experiments} \\
\toprule
\textbf{Study} & \textbf{Task} & \textbf{Sample \& AI Model} & \textbf{Completion Time} & \textbf{Quality} & \textbf{Inequality} \\
\midrule
\endfirsthead

\toprule
\textbf{Study} & \textbf{Task} & \textbf{Sample \& AI Model} & \textbf{Completion Time} & \textbf{Quality} & \textbf{Inequality} \\
\midrule
\endhead

Kreitmeir \& Raschky (2024) & General programming (natural experiment: Italy’s ChatGPT ban) & 36,000 GitHub users; ChatGPT access revoked in Italy & Less-experienced workers faster without AI; experienced worked slightly slower without AI for routine tasks & Less-experienced users’ code quality improved once ChatGPT access was removed & Increase: AI had benefited experienced users more, slightly widening performance gap \\
\addlinespace
Yeverechyahu et al. (2025) & General programming (natural experiment 2019--2022) & 1.1 million commits; Diff-in-diff on Python/Rust (treated) vs R/Haskell (control); GitHub Copilot & Overall number of contributions increases in treatment group; biggest jump in maintenance/iterative work & Focus shifts toward iterative “quality-improving” commits & Not reported \\
\addlinespace
Song et al. (2025) & General programming – project-level code contributions and coordination & 4.9 million repository-month pairs; GitHub Copilot & 5.9\% increase in code contributions; 8\% increase in coordination time & Not measurable effect on code quality & Increase: Core developers gain more than peripheral, widening within-project contribution gap \\
\addlinespace
Hoffmann et al. (2025) & General programming (Copilot introduction, 2022) & 187,489 developers; GitHub Copilot & Not reported; reallocation of time toward coding from management & Not reported & Reduction: lower ability workers benefit more from Copilot access \\
\addlinespace
Daniotti et al. (2025) & General programming in Python (2019--2024) & 80 million commits to open-source Python projects on GitHub & 2.4\% higher output (measured in number of commits) & Not reported & Not reported; higher AI adoption among less-experienced programmers; large country-level differences \\
\bottomrule
\end{longtable}
\end{footnotesize}

\subsubsection{Emerging avenues for future research -  LLM user conversation data \label{subsec:chatbot_data}}

While prior studies have largely relied on variations in AI access or adoption to infer effects on productivity, a new strand of research has begun to examine the content of user interactions with AI systems directly. This shift—from treating AI access as a treatment to analyzing the details of user-AI interactions—offers a more fine-grained understanding of adoption and usage patterns. In a notable contribution, researchers at Anthropic analyze millions of real-time conversations between users and their Claude chatbot, using a privacy-preserving natural language processing technique known as Clio \citep{tamkin2412clio}. Building on this infrastructure, the team constructs an ``Economic Index'' that matches Claude conversations to O*Net tasks and occupations \citep{handa2025economic}.

The findings reveal that coding and writing tasks together account for nearly half of all chatbot usage, while an additional third of interactions span a diverse set of professional tasks, including data analysis, research, and administrative support. While many interactions involve automating relatively simple activities—such as drafting emails, generating templates, or debugging small code snippets—a non-trivial share of usage reflects more iterative, augmentative patterns. Users often refine, adapt, or critique AI outputs through multi-step exchanges, suggesting that conversational agents are increasingly embedded in users’ broader problem-solving processes. This type of usage appears especially prevalent in digitally oriented occupations, where AI tools are being used not just for automation, but also for exploratory reasoning and creative iteration.

Although these results provide some of the most detailed insights to date into how generative AI tools are being used at scale, the approach has important limitations. Access to high-resolution interaction data remains restricted to researchers working at frontier AI labs, limiting independent validation and broader academic scrutiny. Moreover, it is difficult for outside researchers to use LLM user data for experiments, since these data are typically collected and controlled by private companies. While Anthropic has released some of the data publicly, it is only available at highly aggregated levels that constrain its usefulness for empirical analysis. As a result, despite the promise of these datasets for understanding real-world AI usage, their current impact on academic research remains limited. Nonetheless, these studies demonstrate the considerable potential of user-level interaction data to complement existing work on AI adoption and to shed light on how generative tools are reshaping day-to-day work across occupations.

\subsection{Labour demand and supply -- evidence from large-scale digital trace data\label{sec_supply_demand}}

While experiments provide valuable insights into the precise mechanisms through which AI affects workers and firms, they have limitations. By controlling the environment, these studies offer clear causal identification and allow for fine-grained micro-level analysis. However, they primarily focus on direct individual-level productivity effects and do not account for indirect general equilibrium reactions of firms, and how these may affect equilibrium employment and wages. Moreover, they are only valid for the specific task analysed and might not generalise to other occupations. In particular, they are silent on whether the productivity gains will complement or substitute for workers in the broader labour market context (i.e., increase or decrease employment).

To explore these market-level effects, it is necessary to examine other types of studies that leverage larger labour market-level datasets. In the following sections, we turn to research using digital trace data, administrative records and labour surveys to investigate how AI is affecting the labor markets.

Digital trace data refers to the naturally occurring by-products of digital interactions and transactions. As individuals and organizations go about their activities online -- whether writing code on GitHub, posting or responding to job advertisements, freelancing through digital labor platforms or engaging with AI chatbots -- they leave behind a wealth of behavioral data. Unlike experiments or survey-based methods, digital trace data is ``always on'' (\cite{salganik2019bit}): it captures activities as they unfold, often across entire platforms or labor markets. This makes it particularly valuable for studying the adoption and effects of new technologies. Researchers have leveraged such data to conduct rich descriptive analyses and to exploit natural experiments—settings where platform changes or external shocks allow for quasi-experimental identification. While these studies cannot match the rigor of controlled experiments when it comes to establishing causal effects, they can nonetheless uncover important patterns. In particular, they offer valuable insights into real-world behavior at scale, helping to shed light on how technologies are integrated into and affect everyday work practices.

\subsubsection{Online freelancing markets\label{subsec_online_freelancing}}

A growing body of research examines the equilibrium effects of AI on labour markets by leveraging data from online labour markets (OLMs) such as Upwork and Freelancer.com. These platforms provide real-time insights into how AI technologies, including Google Translate and ChatGPT, influence labour demand, job postings, and worker earnings. They offer rich information on both demand (job postings) and supply (worker backgrounds and earnings), span a diverse range of occupations, and reflect a fast-moving labour market with high turnover and little regulation, making them particularly responsive to technological change. However, OLMs cover only a small segment of the overall labour market and largely exclude team- or firm-level dynamics, since most work is organized through short-term individual contracts, which makes it difficult to generalize findings to the broader labour market. By analysing large datasets from these platforms, researchers can explore firms' general equilibrium reactions on the demand side of the labour market as well as how workers adapt their labour supply. 

A common methodology in these studies is the difference-in-differences design, exploiting the staggered adoption of AI tools over time and the differential exposure of various occupations to AI. While they differ in the specific occupations examined (e.g., translators, programmers, writers), these studies reveal common patterns of how AI adoption affects labour demand, particularly for tasks susceptible to automation. 

\cite{Yilmaz2023} investigate the impact of Google Translate on professional translators using OLM data. By comparing different language pairs before and after Google Translate's staggered rollout, they find a significant 13–20 per cent reduction in the number of transactions for "analytic" (regular) translations. In contrast, "interactive" (transcreative) translations that require more creativity remain unaffected. Earnings for translators decline without significant price adjustments, supporting the narrative that AI substitutes human labour in routine tasks but has less impact on complex ones.

Focusing on translation with ChatGPT, \cite{qiao2024ai} examine ChatGPT’s differential impact on freelancers in translation versus web development between May 2022 and October 2023. Exploiting ChatGPT’s November 2022 release as a natural experiment, they find sharp displacement effects in translation: freelancers complete fewer tasks and average earnings decline. In contrast, for web development task completion rises and average earnings increase post-ChatGPT. These divergent outcomes underscore that generative AI does not exert a uniform effect across occupations but rather creates winners and losers, with language-intensive work more exposed to substitution and technical work benefiting from complementarity. In a related analysis, \cite{liu2023generate} find a significant contractions in writing and translation gigs after the introduction of ChatGPT: both the number of jobs and the pool of active freelancers shrank, though demand declined faster than supply, intensifying competition. Transaction volumes and implied earnings fell most sharply in high-volume or lower-quality segments. Yet some freelancers adapted by offering AI-complementary services and benefited from new opportunities. The findings thus illustrate both displacement and adaptation: many workers lost ground, but those who pivoted toward complementary niches could capture gains.

Moving beyond studies that focus solely on LLMs, \cite{hui24} analyse the short-term effects of generative AI on freelancers in online labour markets, focusing on both ChatGPT and image-based models such as DALL·E 2 and Midjourney. Using employment histories from Upwork and a difference-in-differences design, they compare freelancers in occupations most exposed to these tools with less-affected groups. For writing-related services, they find that freelancers in AI-exposed occupations experience a 2 per cent decrease in the number of jobs per month and a 5 per cent decrease in monthly earnings after ChatGPT’s release. Similar adverse effects are observed for design and image services following the launch of image-based AI, reinforcing the robustness of the findings. High-quality workers are also affected, though not always statistically significantly, suggesting that AI’s substitution effect may extend beyond lower-skilled workers and impact a broader range of freelancers.

Using a similar approach, \cite{demirci2025ai} cluster job posts from a leading online labour platform into categories such as writing, software development, engineering, and graphic design, and classify them according to their exposure to AI tools. Exploiting the release of ChatGPT as a shock, they find that demand for automation-prone jobs fell sharply relative to manual-intensive ones, with writing jobs declining by 30 per cent, software and web development by 21 per cent, and engineering by 10 per cent. The release of image-based models (Midjourney, Stable Diffusion, DALL·E 2) also led to double-digit declines in graphic design and 3D modelling jobs. While overall demand contracted in exposed categories, the remaining jobs tended to be more complex and offered higher pay, suggesting a reallocation toward more sophisticated tasks. To validate their results, the authors link variation in job-post declines to Google Trends indices capturing awareness of ChatGPT’s applicability, showing larger declines in clusters with higher search activity. Taken together, the findings point to substantial substitution effects in routine freelance work, coupled with signs of adaptation as employers demand more complex and higher-value tasks.

Purely focusing on design tasks, \cite{lysyakov2023threatened} study the introduction of an automated logo-generation tool in a design crowdsourcing contest platform. They show that many lower-tier designers exited after the AI’s entry, as simple contests were quickly absorbed by the automated system. By contrast, higher-tier designers remained engaged in more complex contests, which were less affected. Although wage effects are not directly measured, the evidence points to an unequal impact: AI displaced less-skilled designers, while those at the top retained opportunities.

Unlike many studies that primarily document the substitution effects of generative AI, \cite{teutloff2025winners} explicitly analyse both the “winners and losers” of AI adoption by examining how demand shifts across substitutable and complementary skills. Drawing on around 3 million job postings from a major freelance platform published between January 2021 and September 2023, they use natural language processing to cluster jobs into skill groups and classify them as substitutable, complementary, or unaffected by AI. Overall freelance demand continued to grow, but with marked reallocation across skills. Demand for substitutable skills like writing and translation decreased by 20-50 per cent relative to the counterfactual trend, especially for short-term jobs. Conversely, demand for certain complementary skills, such as machine learning (which grew by 24 per cent) and AI-powered chatbot development (which nearly tripled), increased significantly. However, across the board of complementary tasks, demand for novice workers decreased. This more granular evidence challenges the notion that AI uniformly destroys jobs or boosts demand across all complementary skills. The study also provides additional evidence challenging the view that low skilled workers benefit the most from AI tools. 

Overall, the studies using online labour market data provide empirical support for the substitution effect of AI on labour demand in routine, automation-prone tasks. The introduction of AI tools like Google Translate, ChatGPT or Midjourney lead to a decline in job postings and earnings for occupations heavily exposed to AI. Conversely, jobs complementary to AI and tasks requiring creativity and complex problem-solving seem to be less affected or may even experience increased demand. These findings suggest that AI induces task reallocation within occupations, with workers shifting towards more complex tasks less susceptible to automation. While the demand for certain skills decreases, there is a simultaneous increase in demand for skills complementary to AI. These findings indicate that, contrary to what some experimental studies have suggested, AI adoption may exacerbate inequality and labor market polarization. Specifically, it seems to benefit workers with skills that complement AI, while displacing those engaged in oftentimes simpler automatable tasks. While these studies offer valuable insights, they primarily capture short-term effects within online freelancing platforms, on selected sub-samples of the workforce, which may not fully represent broader labour market trends. 

\subsubsection{Large online vacancy data\label{subsec:onlinevacancy}}



To assess the labour market effects of AI on a broader, more representative sample of workers than online labor markets, research leverages large-scale online job vacancy data from Lightcast or other providers. These datasets provide detailed information on job postings, including skills and tasks requested by employers, revealing how AI technologies reshape firms' labour demand across industries. Before the release of ChatGPT, \citep{alekseeva21} analyzed U.S. job postings from 2010-2019, finding that AI increased labor demand and wages, particularly for workers with AI skills who earned significant wage premiums. However, \cite{acemogluRestrepo22} used similar data from 2010-2018 to show that while AI-exposed firms increased AI-related hiring, they simultaneously reduced overall hiring. These seemingly contradictory results can be reconciled by acknowledging structural changes in the labor market: while there is augmentation in AI-specific skills, this may come at the cost of other jobs being substituted. This pattern of simultaneous skill premiums and job displacement provides important context for understanding how generative AI affects different worker groups.

Recent work suggest that online job postings are increasingly biased against younger workers \citep{lichtinger25}. Analysing resume and job posting data for a large sample of workers in US companies between 2015 and 2025, the authors identify a significant drop in hiring rates for junior employees after 2023 in those companies that adopted GenAI tools. The drop in hiring was particularly pronounced among mid-tier graduates whereas high- and low-tier graduates did not see a decline in hiring. Senior employment, on the other hand, continued to increase, and no increase in separation rates were reported. These findings seem to echo similar results by \cite{Brynjolfsson25b}, for which the same caveat applies: it is difficult to disentangle the role of AI from broader macroeconomic trends. Shifts in the business cycle, tighter monetary policy, and rising global uncertainty may also have contributed to the observed labor-market patterns.

For China, \cite{ji23} analyse the impact of generative AI on substitution and augmentation effects for different occupational groups. Making use of job vacancy information from a large Chinese job vacancy platform, the authors apply regression discontinuity analysis to understand the impact of the (global) introduction of ChatGPT in November 2022 to determine its impact on the skill characteristics in job vacancy notes. Their analysis covers all new job posting during the period September 2021 and August 2023. They are able to show that over this period job vacancies increasingly shifted to those jobs that required skills complementary to ChatGPT, such as creativity, problem solving or research and development. Other skills such as documentation, image generation or design that can easily be substituted experience significant declines after November 2022 in their database. Such a differential effect of GenAI on the Chinese labour market might have particularly adverse effects for vulnerable groups, such as women and young people as suggested by \cite{wang25} who study the impact of the introduction of automation technologies in China more broadly.

\subsection{AI Adoption and Employment Effects - Evidence from Labour Force Surveys and Administrative Data \label{sec_adoption_employment}}

At a broader level, labour force surveys and administrative data assess AI adoption and its effects on employment, productivity, and wages across regions and industries, with the main advantage being their representativeness of the entire labour market. Administrative data was used to study technology's employment effects  before generative AI emerged. Earlier U.S. studies found that routine-task replacing technologies led to employment reductions and wage decreases of a few percentage points for older workers \cite{kogan23}, while other research showed wage growth increases for workers highly exposed to AI, supporting labor-augmenting effects \cite{Fossen2022}. However, \citep{Autor2024a} demonstrate that while labor-augmenting innovations historically created new occupations, the demand-boosting effects of augmentation have not kept pace with demand-eroding effects of automation in recent decades

Focusing on who adopts generative AI, \cite{humlum2025unequal} link a large, representative Danish worker survey to administrative registers to map ChatGPT use across eleven occupations identified as exposed according to the \cite{Eloundou2023} task-based exposure measures. They find uptake is widespread but uneven: younger and less-experienced workers are more likely to use the tool, while women are about 16 percentage points less likely to have used it for work. Users also tended to earn slightly more even before ChatGPT’s release, suggesting that early adopters were already advantaged in the labour market. Survey respondents see high productivity potential, yet many non-users report employer restrictions and a need for training as key barriers to adoption. These frictions help explain why adoption lags in some workplaces despite perceived benefits. By linking survey responses to administrative records \cite{humlum2025unequal} are able to identify which types of workers adopt ChatGPT -- by earnings, tenure, and demographics -- rather than merely measuring overall adoption rates. The study highlights that, so far, there is no evidence of a broad-based labour market transformation due to generative AI. However, the study points to unequal adoption of the new technology that mirrors -- and may amplify -- existing labour-market disparities.

Building on that descriptive baseline, \cite{humlum25} combine two subsequent adoption surveys from late-2023 and 2024 (covering about 25,000 workers and 7,000 workplaces across the same 11 occupations) with Danish matched employer–employee records to estimate near-term labour-market effects of AI chatbots. They document extensive firm engagement: most employers encourage the use of AI, many deploy in-house models, and training initiatives are common. These measures appear to boost adoption, help narrow usage inequalities, and spur the creation of new job tasks. Despite this, difference-in-differences estimates exploiting variation in employer policies yield tightly estimated null results: there are no detectable effects on earnings or recorded hours in any occupation, with confidence intervals that rule out impacts larger than about 1\%. The authors attribute the muted aggregate effects to modest measured productivity gains (average time savings of only about 3\%) and weak wage pass-through over the study horizon. In other words, adoption has advanced faster than measurable labour-market consequences. This result complements the earlier evidence from \cite{humlum2025unequal}: initial uptake is real but unequal, and even where firms invest, short-run impacts on productivity and wages are negligible. Taken together, the two studies illustrate how combining survey with register data can track both the diffusion of AI tools and their near-term economic impact -- and, so far, point to unequal adoption with limited immediate effects on earnings or employment.

Turning to the U.S. context, \cite{Brynjolfsson25b} provide some of the first evidence of measurable employment effects, focusing specifically on entry-level positions. Using high-frequency administrative payroll data, they document that early-career workers (ages 22–25) in the most AI-exposed occupations experienced a 13 percent relative decline in employment following the spread of generative AI, even after controlling for firm-level shocks. By contrast, employment among more experienced workers in the same jobs, as well as in less exposed occupations, remained stable or continued to grow. The study also shows that adjustment occurs mainly through employment rather than wages, with losses concentrated in occupations where AI is more likely to substitute for rather than complement human labour. Importantly, the authors rule out alternative explanations such as post-COVID labour-market corrections, over-hiring in tech, or remote work dynamics. These findings align with evidence by \cite{lichtinger25} that displacement effects are concentrated among younger, less experienced workers. The results highlight that while aggregate impacts in survey-based studies may appear minimal, granular administrative data reveal early signs of labour-market disruption for the entry level workers.

Overall, evidence from labour force surveys and administrative data suggests that the immediate productivity and employment effects of generative AI are modest compared to the larger impacts documented in experimental settings or studies based on online freelancing data. A likely reason is that labour markets adjust slowly: realizing the full productivity potential of these technologies often requires organizational change, complementary investments, and a reconfiguration of production processes. Moreover, adoption is also uneven, shaped by training needs and firm-level policies that either delay or restrict use. These findings indicate that the diffusion of AI in the wider economy is still at an early stage. Instead of a disruptive transformation, so far, we are observing gradual uptake, institutional frictions, and unequal adoption. This illustrates that the longer-term trajectory of productivity and employment impacts will depend not only on technological capabilities, but also on how firms, workers, and policymakers navigate the new opportunities and challenges of generative AI.

\section{Beyond jobs: The impact of AI on collective performance\label{sec:team-performance}}

The studies reviewed in the previous section focus primarily on individual-level productivity gains from generative AI adoption. However, most work occurs within collaborative contexts—teams, organizations, and broader networks—where workers complement rather than substitute for one another. Understanding AI's impact requires examining how it affects team dynamics, collective intelligence, and organizational coordination. While the empirical evidence remains limited, emerging findings across creativity and innovation, business and marketing, coding, and algorithmic management reveal consistent patterns in how generative AI transforms collaborative work processes. This section examines this growing body of evidence on generative AI's effects on team-based work and collective intelligence.

\subsection{Creativity and innovation\label{subsec:teams_creativity}}

Recent work reveals a paradox in how AI affects collective intelligence: while enhancing individual performance, it simultaneously reduces collective diversity. \cite{lee2024empirical} found that participants using ChatGPT for creative challenges (gift ideas, toy design, household repurposing) produced ideas rated as more creative than those without AI assistance. However, \cite{meincke2025chatgpt} reanalyzed this data focusing on collective-level diversity, finding that AI-aided idea sets had consistently lower diversity—in the extreme case of toy design, only 6\% of ChatGPT-generated ideas were unique (non-overlapping based on semantic similarity) compared to 100\% of human ideas. This pattern extends to creative writing: \cite{doshi2024generative} found that while AI-assisted writers produced stories rated as more creative and enjoyable, these stories were significantly more similar to each other than those written by humans alone, creating a social dilemma, where individual gains come at the cost of collective novelty.

This pattern of AI individual enhancement but collective narrowing extends beyond generative AI. \cite{hao2024ai} analyzed 67.9 million research papers (1980-2024) studying scientists using AI methods, including generative AI but also other types, in their research—from traditional machine learning through recent LLMs. While AI-adopting scientists publish 67.37\% more papers and receive 3.16 times more citations, AI research narrows the field's focus: the "knowledge extent" (the diameter of the vector space covered by research papers in a field) contracts by 4.96\%, and "follow-on engagement" (how frequently papers citing the same work cite each other, indicating emergence of new subfields) reduces by 24.40\%. This suggests AI accelerates work in established data-rich domains rather than catalyzing exploration of new frontiers. Similarly, \cite{riedl2024effects} examined 52,000 chess players using AI analysis tools, finding that while AI feedback improved individual performance, it decreased strategy diversity at the platform level as players converged on similar opening moves.

However, there is also evidence that AI can be used to bridge disciplines. Studies have found that simple autonomous agents can improve creative semantic discovery in human groups, prompting new associations and perspectives that lead to more innovative ideas \cite{Ueshima2024}. Similarly, AI can accelerate scientific discovery by helping researchers connect across disciplines to answer promising questions \cite{Sourati2023}. In other words, while AI may be narrowing the focus within this behavior could also be be used to break down disciplinary silos.

\subsection{Business and marketing}

At the intersection between innovation and marketing, \cite{boussioux2024crowdless} compared crowdsourced worker solutions in generating sustainable circular economy business ideas vs a single researcher using AI (GPT-4) in two ways: non-interactive prompting (fresh GPT-4 sessions for each solution) and interactive prompting (iterative conversations with differentiation instructions). They recruited 148 crowdsourced participants who generated 125 usable solutions, while the researcher generated 730 AI solutions at lower cost (27 hours vs 2,555 hours) and time (5.5 hours vs 2,520 person-hours). The differences in performance were mild but significant in some areas, evaluated by 300 independent raters using 5-point Likert scales. Human crowd solutions exhibited higher novelty both on average (3.51 hours vs 3.23 hours for non-interactive AI, an 8\% advantage) and for top novel outcomes, while showing comparable novelty to interactive AI (3.51 vs 3.47). However, AI solutions scored significantly better on strategic viability (2.7\% higher), environmental value (4.3\% higher), financial value (4.6\% higher), and overall quality (3.0\% higher). The interactive AI approach achieved novelty levels comparable to human crowds while maintaining superior practical value. This study shows that while basic AI usage may decrease novelty, when used strategically by skilled researchers through iterative prompting, AI can lead to solutions of similar novelty with some improvements in performance. 

At least three other studies have also measured productivity benefits of introducing generative AI in teams in marketing, business, and product innovation using field experiments. \cite{dellacqua25} conduct a 2x2 experiment with 776 P\&G participants assigned to four conditions: individuals working alone, human-only teams, individuals with AI, or teams with AI. Participants tackled real business challenges across four problem statements, including helping consumers transition between product forms and motivating trial of new products. Results show that individuals with AI matched the performance of human-only teams, while teams with AI were 12.7\% faster and produced solutions 500+ words longer on average. Critically, teams with AI were 3 times more likely to produce top 10\% quality solutions. The study finds expertise democratization, employees with limited product development experience, AI enabled individuals to achieve performance levels comparable to experienced teams. Interestingly, without AI, R\&D and Commercial professionals produce distinctly technical vs. commercial ideas, whereas with AI their output distributions converge. This suggests that AI can help reduce disciplinary silos, albeit at the cost of narrowing diversity.

\cite{li24} studied four groups: human-only teams, teams with one shared AI, teams where each member had their own AI, and individuals working alone with AI (in a separate experiment). Participants completed two professional tasks—creating articles on college job hunting and developing a digital transformation strategy for a retail chain. Adding AI to teams improved solution quality, novelty, and usefulness by 2-4 percentage points compared to human-only teams. Teams with multiple AIs didn't perform better than teams with one AI. The most effective approach was when one or few team members used the AI deeply rather than distributed usage. Individuals working alone with AI that did not spend enough time analysing and checking the LLMs output, showed dramatically poor initial performance, scoring 38 to 44\% lower on quality, novelty, and usefulness compared to human-only teams.  When controlling for time spent on the task, solo AI users matched human-only teams but did not reach AI-assisted team performance. This suggests AI helps individuals work faster, but overseeing AI models and curating output is crucial. More qualified teams benefited significantly more from AI—teams with higher individual cognitive ability, stronger relationships, and larger sizes were better at leveraging multiple AI tools effectively, echoing some of the theoretical findings in the literature on collective intelligence review in section \ref{sec:collective_intelligence}. 

\cite{ju2025collaborating} conducted a controlled online experiment where participants designed ads for a large think tank, split between human-human teams and human-AI teams. Human-AI teams produced 60\% to 73\% more ads and sent 23\% fewer social messages than human-human teams. Overall, ads created by human-AI teams performed similarly to human-human teams in field tests, with comparable click-through rates. One interesting finding was that prompting different personas into AI and pairing with different human personalities significantly affected team performance. Field experiments on GenAI in business and knowledge work remain scarce. The one relevant study we identified, \cite{dillon2025shifting} discussed previously, suggests some findings translate. Mainly the productivity gains and reductions in written communication (less emails), yet the organizational restructuring appears limited, with meeting patterns unchanged. Perhaps organizational change requires more flexible or dynamic work environments, such as coding and collaborative software development.

\subsection{Coding}
AI is transforming how developers organize their work on collaborative projects. \cite{Hoffmann2025NatureOfWork} exploit an internal GitHub ranking system where developers below a threshold received free Copilot access, applying regression discontinuity design to 187,489 developers observed weekly from July 2022 through July 2024. Coding activities increase by 5.4 percentage points, but most interestingly project management activities drop by 10 percentage points, demonstrating that developers shift from managerial coordination tasks toward core technical work. They also find developers engaging with 15 additional repositories while collaborator counts drop 79.3\% as work becomes more autonomous.
\cite{Yeverechyahu24} leverage Copilot's selective language support, comparing Python and Rust packages (supported) with R and Haskell packages (unsupported) using difference-in-differences analysis of 1.1 million commits from October 2019 to December 2022. Beyond increasing commits by 37-54\%, they distinguish between capability innovation (commits adding new functions) and iterative innovation (commits without new functions, focused on maintenance and refinement). While both types increase, iterative innovation grows substantially more: 39.1\% versus 26.5\% for capability innovation, with the absolute effect 7.5 times larger.
These studies reveal the individual-collective creativity paradox in coding. At the individual level, developers using AI explore more repositories and work more autonomously, bypassing collaborative friction. However, at the project level, they increasingly focus on iterative improvements rather than breakthrough capabilities. There is, however, also the analogous of breaking silos, \cite{daniotti2025using} find that software developers using AI are more likely to use new libraries.

\subsection{Algorithmic management}
A particular case of having generative AI as a teammate occurs when AI assumes managerial functions, assigning tasks and coordinating work among team members. As managers, LLMs can initiate interactions between workers, suggest collaborations, allocate responsibilities, and provide guidance on task execution. This represents a specialized application of algorithmic management—the use of software to partially or fully automate tasks traditionally carried out by human managers \cite{milanez2025algorithmic}. An extensive review of algorithmic management lies outside the scope of this work. However, we note that algorithmic management predates generative AI and has been extensively studied across digital labor platforms and traditional workplaces. The broader literature documents significant negative impacts on worker well-being, including increased surveillance, work intensification, job insecurity, loss of autonomy, and decreased job satisfaction \cite{Kellogg2020, Mohlmann2021, niehaus22, baiocco22}. Cross-national evidence reveals widespread adoption—reaching 90\% in the United States and 79\% across European countries—but also substantial worker concerns, with 23\% reporting negatively affected job satisfaction and 29\% feeling uncomfortable with monitoring practices \cite{milanez2025algorithmic, braten2019monitoring}. While some negative effects may be mitigated through proper implementation \cite{Kusk2022}, the literature reveals a pattern where potential managerial efficiency gains often come at the cost of worker autonomy and well-being.

Given this context of documented concerns with traditional algorithmic management, how do workers respond specifically to generative AI managers? Early evidence presents a nuanced picture. \cite{dong2024} conducted an MTurk field experiment where participants experienced no significant differences between AI and human management across performance, motivation, fairness perceptions, and future commitment measures. However, the same study's survey revealed strong disapproval of replacing human managers with AI, illustrating an attitude-behavior gap between hypothetical preferences and actual experiences. Two mechanisms help explain negative attitudes toward AI management. First, people consistently perceive lower social status when managed by algorithms compared to human managers, as this signals that job tasks lack complexity \cite{jago2024}. Across five preregistered studies including live interactions with a GPT-3-based manager delivering instructions and feedback, participants reported lower status and more negative emotions under algorithmic management. Second, an "empathy penalty" emerges where people rate identical supportive messages as less empathetic when labeled as coming from AI rather than humans \cite{liu2025illusion, perry2023ai}. These findings suggest that while workers may adapt to AI management in practice, status concerns and empathy deficits represent persistent challenges for generative AI adoption in managerial roles.

\subsection{Take Aways\label{subsec:teamsTakeAways}}
While the literature on generative AI and collective intelligence is still growing, there are some consistent findings regarding work reorganization and team productivity. Studies show AI improves performance across diverse tasks—from business strategy development and creative writing to academic research and software development \cite{dellacqua25,li24,ju2025collaborating,hao2024ai,Yeverechyahu24}. However, teams composed of a human and an AI can show worse performance when AI is not supervised correctly \cite{li24}, or less novelty when AI is not used iteratively \cite{boussioux2024crowdless}. The O-ring theory provides a framework for understanding these patterns: when AI is not used properly, it becomes the weakest link in the production chain, degrading overall performance. AI also systematically reduces coordination overhead, developers spend less time on management tasks and shift toward core technical work \cite{Hoffmann2025NatureOfWork}, while human-AI teams share fewer social messages  \cite{ju2025collaborating}.

In terms AI assuming managerial roles, field evidence does not show any measure difference in performance. However, workers dislike being managed by AI, perhaps due to perceived status loss and an “empathy penalty”. Given documented autonomy and well-being risks from algorithmic management, this warrants caution and further study.

The literature also reveals an individual-collective innovation paradox across domains. While AI enhances individual performance in creative tasks \cite{lee2024empirical}, academic research \cite{hao2024ai}, and strategic games \cite{riedl2024effects}, it simultaneously reduces collective diversity—from unique ideas in creative tasks \cite{meincke2025chatgpt} to knowledge extent in academic fields \cite{hao2024ai} and strategy diversity in chess \cite{riedl2024effects}. This narrowing may be offset by increased volume from productivity gains and AI's potential for interdisciplinary bridging \cite{Ueshima2024,Sourati2023,daniotti2025using}. Whether AI will ultimately narrow collective exploration or foster groundbreaking interdisciplinary discoveries remains an open question that deserves further empirical investigation.

\section{Conclusion}

\paragraph{Summary.}

Despite the widespread adoption of AI, measurable effects on job opportunities, wages or inequality remain limited. As the latest wave of GenAI is still in its early stages, understanding the long-term effects of AI on employment is difficult. However, there is a growing literature that can help us understand the potential effects. By synthesizing theoretical frameworks, exposure measures, and empirical evidence across multiple methodological approaches, our review identifies patterns that help explain AI’s heterogeneous effects while highlighting critical gaps in our understanding. A key contribution of our analysis is the classification of task complexity along four dimensions: knowledge requirements, clarity of goal, interdependence, and context requirements. This framework helps explain why AI’s effects vary so dramatically across different types of work.

Understanding AI's labor market impact requires increasingly granular theoretical frameworks. The literature has evolved from aggregate production functions treating labor as homogeneous, to models distinguishing high- and low-skilled workers, to occupation-level heterogeneity, and finally to task-based frameworks with expertise distinctions. This shift represents more than methodological refinement—it reveals fundamentally different mechanisms through which technology affects work. Each level of granularity explained patterns invisible at higher aggregation levels. Where aggregate models predicted uniform productivity gains, skill-based models revealed widening wage gaps. Where occupation-level analysis suggested certain jobs were safe from automation, task-based frameworks showed how AI could transform work within occupations. Most recently, the expertise framework shows that the same AI application can both democratize access to expert knowledge and create new specialized roles. This dual effect helps explain seemingly contradictory empirical findings of inequality reduction and inequality amplification.



There are three prominent methods for measuring workers' exposure to automation: expert surveys, patent data analysis, and AI-generated measures. While these methods often yield different results, recent metrics generally agree that high-wage and cognitive occupations are more exposed to AI. However, whether this exposure represents substitution or complementarity is not always clear. Drawing firm conclusions on the future of work is also difficult given that AI is just one of many technologies impacting the labour market and that labour mobility plays an important role. Moreover, some assessments of the impact of AI on particular labour market groups suffer from the short time span over which the evidence has been collected, which often does not allow to separate the impact of technological change from other, macro-economic factors such as the recent rise in economic uncertainty. Overall, exposure itself is not a sufficient statistic to predict wages or employment opportunities across occupations exposed to AI..


The complexity of the labour market impact of AI becomes even more evident when considering the experimental evidence. Controlled experiments reveal AI’s mixed effects on productivity and employment. While AI tools can significantly enhance productivity in simple tasks, they can also lead to poorer outcomes in more complex tasks, including diminished performance and worker displacement. Online labour market studies also point to substitution effects. Given that several firms use freelancers for short tasks, this may be a substitution of freelancers for in-house labour. There is also some evidence of an increase in demand for freelancers in niche AI jobs, such as chatbot development. Broader job vacancy data reveals a reallocation of demand across skill types, with Chinese firms shifting job postings toward complementary skills like creativity and problem-solving while reducing demand for easily substitutable skills like documentation and design. Administrative data from the U.S. shows particularly concerning patterns for young workers, with companies adopting AI reducing hiring of junior employees (ages 22-25) by 13\%, suggesting that AI may be creating significant barriers to entry-level employment, though these effects vary considerably across countries—Denmark, for instance, shows negligible employment impacts of less than 1\%.


At the same time, AI adoption raises important questions about team dynamics and the reorganization of labour within firms. As AI tools take on more collaborative roles within teams, the boundaries between human and AI contributions may blur. While these innovations promise to enhance productivity and streamline decision-making, they also risk creating over-reliance on AI, which could undermine human judgment and the collective intelligence that typically drives successful teamwork. Moreover, GenAI tools show a "regression to the mean" \citep{ernst25}, which reduces diversity of output, often an essential ingredient for successful innovations. This creates a trade-off between individual quality and variability, which, for instance, shows up in grand applications for research funds \citep{dellacqua23}. Too intensive a use of GenAI might reduce the overall variation of grant applications that a competitive bidding process might get, thereby limiting the potential for radical innovation.

\paragraph{Outlook and future research.}

Looking ahead, several research questions remain. Measuring AI’s impact on wages and employment at regional and industry levels continues to be challenging, particularly due to endogeneity issues and the relatively early stage of AI adoption. While the effects may be positive for certain sectors, they are likely to be uneven and difficult to identify in the short term. Future research will need to address these complexities by developing more consistent and reliable exposure measures, examining occupational mobility disruptions, and conducting experiments across a wider range of both simple and complex tasks. Additionally, leveraging data from broader labour market- and company-level surveys -- including longitudinal studies across industries and regions -- will be crucial to understanding how AI is being integrated and how it reshapes both in-house labour demand, team dynamics and organisational restructuring. Moreover, there is still a gap in the literature when it comes to understanding how AI may affect labour markets in developing countries. 

Recent research also points to indirect adverse effects of the increased use of GenAI tools. Indeed, \cite{Kosmyna25} identify a possible reduction in the cognitive capacity of participants in an experiment on LLM-assisted essay writing. If these findings are confirmed, an intensive use of GenAI tools in daily work could reduce the cognitive breadth of users with potential implications for their capacity to carry out their assigned tasks appropriately.

Moreover, few of the reviewed research provided detailed policy recommendations. Most of the papers cited here would emphasise the importance of digital skills and the development of skills complementary to GenAI tools, notably interpersonal skills. Recent research, however, put up some doubt regarding the importance of digital skills. \cite{richiardi25} do not find any evidence for their importance in the previous wave of digitalisation in Europe, pointing out that digital skills might act more as a signaling device rather than an actually productive skill that would prevent job loss.

\newpage

\bibliographystyle{ilostyle}
\bibliography{main}

\section{Appendix}

\subsection{Additional AI Exposure Measures}

\begin{figure}[H]
    \centering
    \caption{AI Exposure Scores by Broad Occupation Groups (China)\label{fig:china_scores}}
    \includegraphics[width=0.7\linewidth]{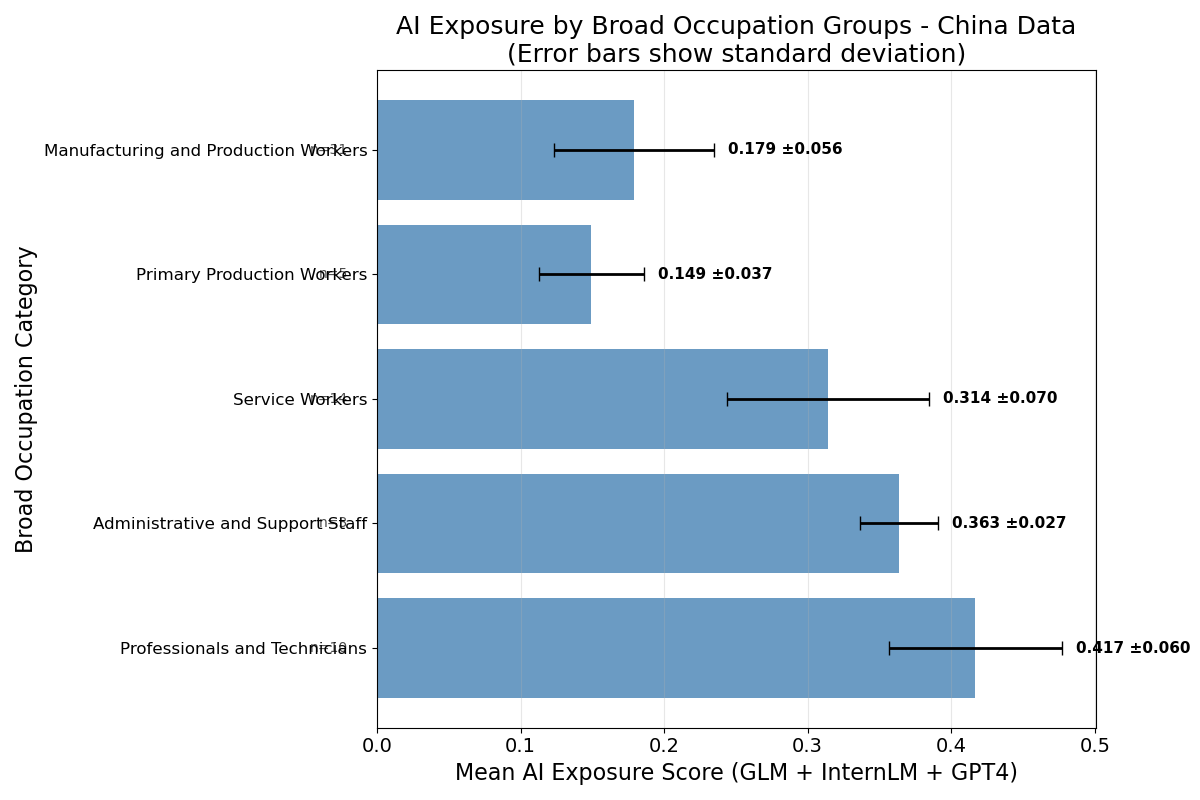}
    
    \pbox{15cm}{\textit{Notes:} AI exposure scores from \citet{chen25} aggregated by broad occupation categories using the Chinese Standard Classification of Occupations. Bars represent mean exposure scores across occupations within each category, with error bars indicating standard deviations. Exposure scores are based on a composite measure combining GLM, InternLM, and GPT-4 assessments. Sample sizes (n) for each category are shown on the left.}
\end{figure}

\begin{figure}[H]
    \centering
    \caption{AI Exposure Scores by ISCO Sub-Major Occupation Groups (United States)\label{fig:isco_scores}}
    \includegraphics[width=0.8\linewidth]{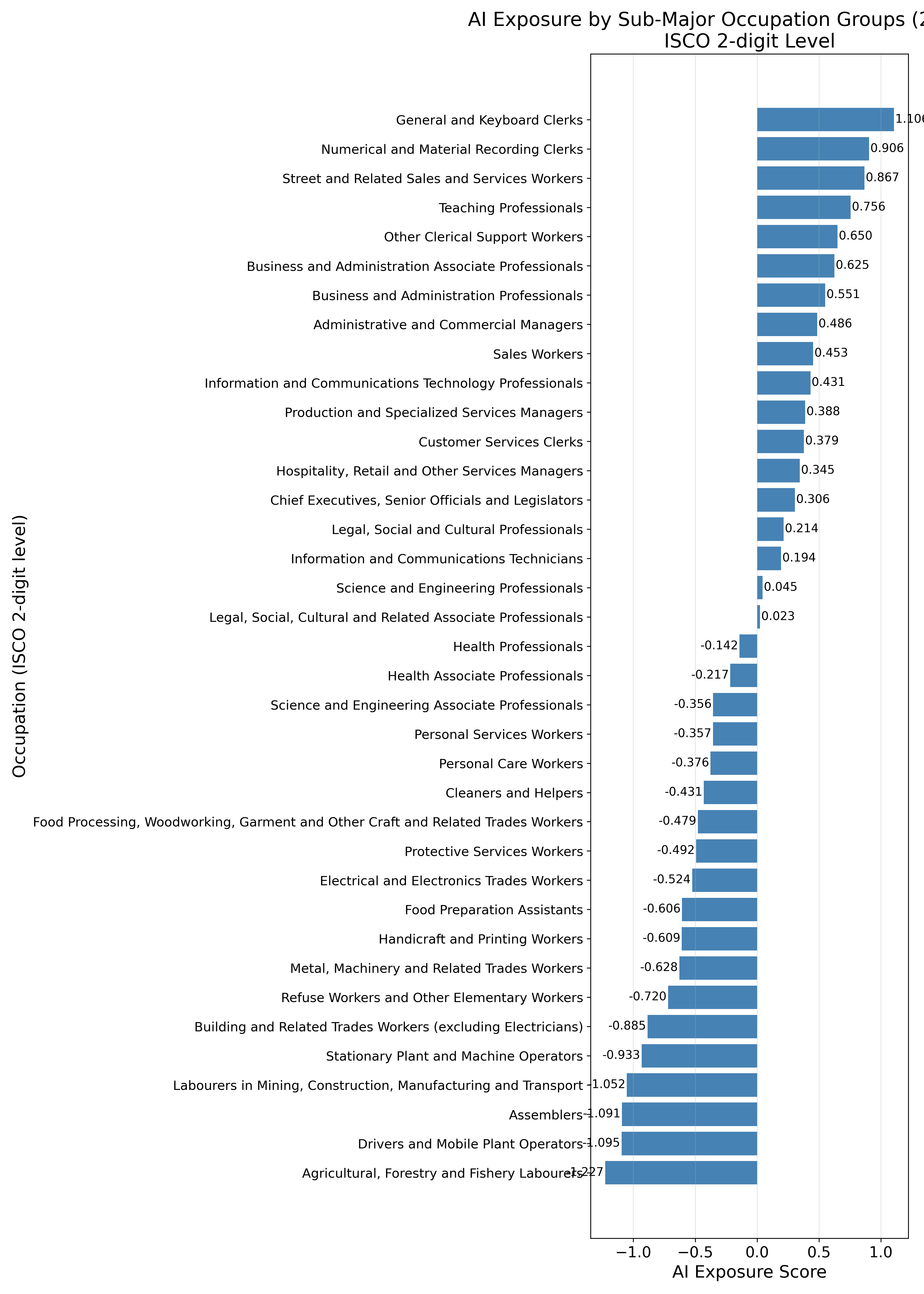}
    
    \pbox{15cm}{\textit{Notes:} AI exposure scores from \citet{lewandowski25} aggregated at ISCO 2-digit level for the United States, 2012-2014. Higher scores indicate greater exposure to AI automation. ISCO 2-digit categories represent sub-major occupation groups providing detailed occupational breakdowns (e.g., Science and Engineering Professionals, Health Professionals, Teaching Professionals).}
\end{figure}

This section presents complementary AI exposure measures from two recent studies that provide broader occupational perspectives. Figure \ref{fig:isco_scores} displays exposure measures from \citet{lewandowski25} using the International Standard Classification of Occupations (ISCO) framework for the United States, while Figure \ref{fig:china_scores} shows exposure measures from \citet{chen25} for China using the Chinese Standard Classification of Occupations. While these measures are more aggregated than our main analysis, they offer valuable comparative insights into AI exposure patterns across different occupational structures and national contexts.

\newpage


\subsection{Classification Rubric - Task Complexity}\label{appendix:task-complexity}

We determine task complexity by evaluating tasks against four dimensions, each assessed as a binary yes/no classification. A task is considered complex if it meets the high-complexity criteria (scores "yes") in at least three of the four dimensions. Tasks that meet fewer than three criteria are classified as simple.

\begin{itemize}
    \item Dimension 1: Knowledge Requirements

A task scores "yes" for knowledge complexity when it requires undergraduate or graduate-level domain expertise beyond the first few years of study or professional experience. This includes tasks that require deep specialization, advanced theoretical understanding, or cutting-edge knowledge in a particular field. Tasks requiring only high school knowledge, basic undergraduate concepts, or entry-level professional skills score "no" for this dimension.

\item Dimension 2: Goal Definition and Success Criteria

A task scores "yes" for goal complexity when it has ambiguous objectives, unclear success criteria, or open-ended exploration requirements. These tasks often involve subjective evaluation standards, multiple valid approaches, or discovery-oriented work where the end state cannot be precisely defined beforehand. Tasks with clear, measurable objectives and unambiguous success criteria score "no" for this dimension.

\item Dimension 3: Task Interdependence 

A task scores "yes" for interdependence complexity when it involves multiple components with critical dependencies, following an O-ring theory model where failure of any element causes system-wide failure. These tasks require complex coordination across many elements and exhibit non-linear relationships between components. Tasks that can be completed independently or with minimal dependencies on other work score "no" for this dimension.

\item Dimension 4: Resources and Context Requirements

A task scores "yes" for resource complexity when it requires extensive materials, multiple data sources, specialized tools, or significant contextual information. These tasks often involve large context windows, hard-to-access resources, or comprehensive background information that must be considered for successful completion. Tasks requiring no external resources or just simple, readily available materials with minimal context requirements score "no" for this dimension.

\end{itemize}

The following table shows examples of how studies were classified according to task complexity using our four-dimensional rubric. Tasks scoring "yes" (1) on at least three of four dimensions are classified as complex, while those scoring fewer than three are classified as simple. D1: Knowledge requirements; D2: Goal definition and success criteria; D3: Task interdependence; D4: Resources and context requirements.
\begin{footnotesize}
\rowcolors{2}{gray!15}{white}
\begin{longtable}{L{2.2cm} L{3.5cm} L{1.2cm} L{1.2cm} L{1.2cm} L{1.2cm} L{0.8cm} L{1.2cm}}
\caption{Examples of Task Complexity Classification}\label{tab:task-complexity} \\
\toprule
\textbf{Study} & \textbf{Task Description} & \textbf{D1: Knowledge} & \textbf{D2: Goal Definition} & \textbf{D3: Interdependence} & \textbf{D4: Resources and Context} & \textbf{Sum} & \textbf{Label} \\
\midrule
\endfirsthead

\toprule
\textbf{Study} & \textbf{Task Description} & \textbf{D1: Knowledge} & \textbf{D2: Goal Definition} & \textbf{D3: Interdependence} & \textbf{D4: Resources and Context} & \textbf{Sum} & \textbf{Label} \\
\midrule
\endhead

\multicolumn{8}{l}{\textbf{Simple Tasks (Score 0 to 2)}} \\
\midrule

Noy \& Zhang (2023) & Professional writing: press releases, reports, emails & 1 & 0 & 0 & 0 & 1 & Simple \\
\addlinespace

Brynjolfsson et al. (2025) & Customer support: live chat technical assistance & 0 & 0 & 0 & 1 & 1 & Simple \\
\addlinespace

Peng et al. (2023) & Programming: implement HTTP server in JavaScript & 1 & 0 & 0 & 0 & 1 & Simple \\
\addlinespace

Dillon et al. (2025) & General knowledge work: emails, documents, meetings & 0 & 0 & 0 & 0 & 0 & Simple \\
\addlinespace

Choi et al. (2024) & Legal writing: contracts, complaints, memos & 1 & 0 & 0 & 1 & 2 & Simple \\
\addlinespace

Kim \& Moon (2024) & Mixed tasks: reading, writing, math problem-solving & 0 & 0 & 0 & 0 & 0 & Simple \\
\addlinespace

\midrule
\multicolumn{8}{l}{\textbf{Complex Tasks (Score 3 or 4)}} \\
\midrule

Dell'Acqua et al. (2023) & Management consulting: product innovation and strategy & 1 & 1 & 1 & 0 & 3 & Complex \\
\addlinespace

Usdan et al. (2024) & Academic writing: scientific analysis and policy proposals & 1 & 0 & 1 & 1 & 3 & Complex \\
\addlinespace

Otis et al. (2023) & Entrepreneurship: business strategy for small enterprises & 0 & 1 & 1 & 1 & 3 & Complex \\

\bottomrule
\label{tab:example_class}
\end{longtable}
\end{footnotesize}

















\textcopyright{} 2025 The Authors. All rights reserved. Please do not reproduce, distribute, or adapt this
work without the permission of the authors. Citation is permitted. This version has not
undergone peer review.

\end{document}